\documentclass[10pt, a4paper, conference]{IEEEtran}

\usepackage[left=1.2cm, right=1.2cm, top=1.8cm, bottom=2.5cm]{geometry} 
\usepackage{epsfig}
\usepackage{epstopdf}
\usepackage{times}
\usepackage{array}
\usepackage{amssymb}
\usepackage{color}
\usepackage{subfig}
\usepackage{amsmath}
\usepackage{amsthm}
\usepackage{multirow}
\usepackage{framed}
\usepackage{color}
\usepackage{cancel}
\usepackage{cite}
\usepackage{bm}
\usepackage{verbatim}
\usepackage{footnote}
\usepackage{multirow}
\usepackage[table,xcdraw]{xcolor}
\usepackage{booktabs}
\usepackage[capitalise]{cleveref}
\usepackage{algorithm}
\usepackage{algorithmic}
\usepackage{fancyhdr}
\usepackage[normalem]{ulem}
\usepackage{slashbox}

\newtheorem{definition}{\textbf{Definition}}
\newtheorem{proposition}{\textbf{Proposition}}

\newtheorem{theorem}{\textbf{Theorem}}
\newtheorem{corollary}{\textbf{Corollary}}

\newif\iffull
\fulltrue

\begin{document}
\title{Elastic Multi-resource Network Slicing:  \\
Can Protection Lead to Improved Performance? }
\author{\IEEEauthorblockN{Jiaxiao Zheng,
Gustavo de Veciana
}
\IEEEauthorblockA{Dept. of Electrical and Computer Engineering, The University of Texas at Austin, TX}
\IEEEauthorblockA{gustavo@ece.utexas.edu}
}

\pagenumbering{arabic}
\thispagestyle{plain}
\pagestyle{fancy}
\maketitle

\begin{abstract}
In order to meet the performance/privacy requirements of future
data-intensive mobile applications, e.g., self-driving cars,
mobile data analytics, and AR/VR, service providers 
are expected to draw on shared storage/computation/connectivity resources
at the network ``edge". To be cost-effective, a key functional requirement
for such infrastructure is enabling the sharing of heterogeneous resources 
amongst tenants/service providers supporting spatially varying 
and dynamic user demands.
This paper proposes a resource allocation criterion, namely, Share Constrained 
Slicing (SCS), for slices allocated predefined shares of the network's resources,
which extends traditional $\alpha-$fairness criterion, 
by striking a balance among inter- and intra-slice 
fairness vs. overall efficiency.
We show that SCS has several desirable properties including slice-level 
\emph{protection}, 
\emph{envyfreeness}, 
and \emph{load driven elasticity}.
In practice, mobile users' dynamics 
could make the cost of implementing SCS high, so we discuss the 
feasibility of using a simpler (dynamically) weighted max-min as a surrogate resource
allocation scheme. For a setting with stochastic loads and elastic user requirements,
we establish a sufficient condition for the stability of the associated coupled network system.
Finally, and perhaps surprisingly, we show via extensive simulations that 
while SCS (and/or the surrogate weighted max-min allocation) provides inter-slice protection, 
they can achieve improved job delay and/or perceived throughput, as compared to
other weighted max-min based allocation schemes whose intra-slice weight allocation
is not share-constrained, e.g., traditional max-min or discriminatory processor sharing.
\end{abstract}

\maketitle

\section{Introduction}
Next generation networks face the 
challenge of supporting data-intensive services and 
applications, such as self-driving cars, infotainment,
augmented/virtual reality \cite{Sat17}, 
Internet of things \cite{BMZ12,Sat17},
and mobile data analytics \cite{CSK18, AhA16}.
In order to accommodate the performance and privacy requirements of 
such services/applications, providers 
are expected to draw on shared storage/computation/connectivity resources
at the network ``edge".
Such network systems can take advantage of 
Software-Defined Networking and Network Function Virtualization technologies
to provision slices of shared heterogeneous resources
 and network functions  which are customized to service providers'/tenants' requirements. 

The ability to support slice-based provisioning is central to enabling 
service providers to take control of managing performance 
of their own dynamic and mobile user populations. This also improves
the scalability by reducing the complexity of performance management on 
multi-service platforms. 
The ability to efficiently share network/compute resources 
is also key to reducing the cost of deploying such services. 
By contrast with today's cloud computing platforms, our focus in this paper is
on provisioning slices of edge resources  
to meet mobile users/devices requirements. In general, shared edge 
resources will have smaller overall capacity resulting 
in reduced statistical multiplexing and making efficiency critical. 
Perhaps similarly to cloud computing platforms, providers/tenants 
will want to make long-term provisioning commitments 
enabling predictable costs and resource availability, yet benefit,
when possible, of elastic resource allocations aligned with 
spatial variations in their mobile workloads but not at the expense of
other slices. 
Thus a particularly desirable feature is to enable slice-level provisioning 
agreements  which achieve inter-slice protection, load-driven elasticity 
and network efficiency.

These challenges distinguish our work from previous research in areas including
engineering, computer science and economics.
The standard framework used in communication
networks is utility maximization (see e.g., \cite{SrY13} and references therein),
which has led to the design of several transport and scheduling
mechanisms and criteria, e.g., the widely discussed proportional fairness.
When considering dynamic/stochastic networks, e.g., \cite{BoM01}, \cite{DLK01},
researchers have studied networks where users are allocated resources based
on utility maximization and studied requirements for 
network stability for `elastic' user demands, e.g., file transfers. 
This body of work emphasizes user-level resource allocations,
without specifically accounting for 
interactions among slices.
Thus, it does not directly address the requirements of network slicing. 

{\color{black}
Instead in this paper, we propose a novel approach,
namely, \emph{Share Constrained Slicing (SCS)}, wherein each slice
is assigned a share of the overall resources, and in turn, distributes
its share among its users. 
Then the user level resource allocation
is determined by maximizing a sharing criterion.
When SCS is applied to a setting
where each user only demands one resource,}
for example, slices sharing wireless resources
in cellular networks \cite{CBD16, CBV17, ZCV18},
it can be viewed as a Fisher market
where agents (slices), which are share (budget) constrained,
bid on network resources, see, e.g., \cite{NRT07}, and for applications \cite{Ban02,FLZ09,CBD16}.
However, those works do not deal with settings where users
require heterogeneous resources, and how to orchestrate 
slice-level interactions on different resources is not clear yet.

When it comes to sharing on heterogeneous resources,
a simple solution is static partitioning of all resources 
according to a service-level agreement, see, e.g., \cite{GuA13}. 
It offers each slice a guaranteed allocation of the 
network resources thus 
in principle provides ideal protection among slices. 
However, it falls short from the perspective of providing load-driven
elasticity to a slice's users, possibly resulting in either resource
under-utilization or over-booking.
Other natural approaches
include full sharing \cite{AAA06}, where users from all slices are served based on
some prioritizing discipline without prior resource reservation.
Such schemes may not achieve
slice-level protection and are vulnerable
to surging user traffics across slices. 

Additionally, many resource sharing schemes have been proposed 
for cluster computing where heterogeneous resources are involved, 
including Dominant Resource Fairness (DRF)\cite{GZH11}, 
Competitive Equilibrium from Equal Income (CEEI) \cite{Mou04}\cite{Var73}\cite{You95},
Bottleneck Max Fairness (BMF) \cite{BoR15}, etc.
These allocation schemes are usually based on modelling joint
resource demands of individual users, but lack of the notion
of slicing, thus it is not clear how to incorporate the need to enable 
slice-level long-term commitments. In these works, inter-slice protection
and elasticity of allocations have not been characterized. Furthermore,
most of these works are developed under the assumption that users
are sharing a centralized pool of resources. In this 
paper we focus on a settings where resources are distributed,
and mobile users are restricted to be served by proximal edge resources.

\emph{Contributions:} The novelty of our proposed approach lies in maintaining slice-level 
\emph{long-term commitments} defined by a service-level agreement, while 
enabling user-level resource
provisioning which is driven by dynamic user loads. 
We consider a model where users possibly require heterogeneous 
resources in different proportions,
and the processing rate of a user scales linearly 
in the amount of resources it is allocated.
Such a model captures tasks/services which 
speeds up in the allocated resources, which is discussed
further in the sequel.

We show that SCS can capture inter- and intra-slice fairness 
separately.
When viewed as a resource sharing criterion,
SCS is shown to satisfy a set of axiomatically desirable properties
akin to those in \cite{LKC10}, and can be interpreted as achieving 
a tunable trade-off among inter-slice fairness (which can be 
seen as a proxy of protection), intra-slice fairness,
and overall utilization.
Fairness is connected to load-driven elasticity 
through share constrained weight allocation.
The merits of SCS are demonstrated in both static and dynamic
settings. In static settings, we prove a
set of desirable properties of SCS as a sharing criterion, including slice-level
\emph{protection} and \emph{envyfreeness}, and we demonstrate the feasibility
of using a simpler (dynamically) weighted max-min as a surrogate resource
allocation scheme for the cases where the cost of implementing SCS
is excessive.
In a dynamic settings, we consider the elastic traffic model where each user carries
a fixed workload, and leaves the system once the work is processed.
We model such system as a stochastic queuing network,
and establish its stability condition. 

Finally, and perhaps surprisingly, we show
via extensive simulations that while SCS provides inter-slice protection,
it can also achieve improved average job delay and/or perceived throughput, 
as compared with multiple variations of traditional (weighted) max-min fair allocations
but without share-constrained weight allocation.
We provide a heuristic explanation of such improvement that SCS can 
separate the busy-periods of different slices, thus reduces inter-slice 
contention, and validate the explanation through simulations.

\iffull
\emph{Paper organization:} This paper is organized as follows. In Section \ref{sec:fairness}, we 
establish our model for network slicing on heterogeneous shared resources,
and characterize SCS as satisfying several axiomatically desirable 
properties for fairness criterion. Then, in Section \ref{sec:static}, 
several properties of SCS involving slice-level utility are discussed,
including protection, envyfreeness, and the feasibility of using a simpler weighted
max-min as a surrogate resource allocation scheme. 
In Section \ref{sec:elastic},
for a setting with stochastic loads
and elastic user traffic, we establish the stability condition when 
SCS is applied as the service discipline. Finally, in Section \ref{sec:simulations},
extensive simulations are conducted for both simple settings where 
only one resource is shared amongst slices, and complex settings which
resembles a realistic edge computing scenario, to demonstrate
the surprising result that SCS
and/or the surrogate weighted max-min allocation can achieve improved average 
job delay and/or perceived throughput while provides inter-slice protection.
\else
\fi

\section{Resource sharing in network slicing}
\label{sec:fairness}
In this section we will briefly introduce the overall framework for
resource allocation to network slices,
namely, Share Constrained Slicing (SCS)
where each slice manages a possibly dynamic set of users. 
Specifically, we will consider resource allocation driven 
by the maximization of an objective function geared at
achieving a trade-off between overall efficiency and fairness \cite{LKC10}. 

To begin, we consider the set of active users on each slice
to be fixed.
Let us denote the set of slices by $\mathcal{V}$, 
the set of resources by $\mathcal{R}$, 
each with a capacity normalized to 1.
Each slice $v$ supports a set of user classes,
denoted by $\mathcal{C}^v$, and the total set of user
classes is defined as $\mathcal{C} := \cup_{v\in\mathcal{V}}
\mathcal{C}^v$.
For simplicity, we let $v(c)$ denote the slice which supports
class $c$.
We let $\mathcal{U}_c$ denote the set of users of class $c$,
and the users on slice $v$ is denoted by
$\mathcal{U}^v := \cup_{c\in\mathcal{C}^v} \mathcal{U}_c$.
Also, the overall set of users is $\mathcal{U} := \cup_{v\in\mathcal{V}}
\mathcal{U}^v$. For each user, possibly heterogeneous resources are required
to achieve certain processing rate. Let us denote the processing
rate seen by user $u$ by $\lambda_u$.
We also define the resource demand vector of user class
$c$ as $\mathbf{d}_c := (d^r_c : r \in \mathcal{R})$,
where $d^r_c$ is the fraction of resource $r$ required
by user $u\in\mathcal{U}_c$ for a unit processing rate,
i.e., to achieve $\lambda_u = 1$, we need to allocate 
fraction $d^1_c$ of the total
amount of resource $1$ to $u$, $d^2_c$ of the total amount
of resource $2$ to $u$, and so on. 
If a user class $c$ does not use a given resource $r$ then $d^r_c = 0$.
Note that if two slices support users with the same requirements,
we will distinguish them by defining two distinct user classes one 
for each slice.
In other words, it is possible to have more than one user classes with exactly the same
$\mathbf{d}_c$.
Also, we let $\mathcal{R}_c$ denote the set of resources
required by users of class $c$,
and in turn, let the set $\mathcal{C}_r$ denote user classes using resource $r$.
Among $\mathcal{C}_r$, the set of classes on slice $v$ is
$\mathcal{C}^v_r := \mathcal{C}_r \cap \mathcal{C}^v$.
The number of active users of class $c$ at time $t$ is denoted by a random
variable $N_c(t)$, and that on slice $v$
by $N^v(t)$.
$N^v(t) = \sum_{c\in\mathcal{C}^v} N_c(t)$.
Realizations of these are denoted by lower case variables
$n_c$ and $n^v$, respectively. 

This model captures the
services/applications where tasks 
speed up with more allocated resources, e.g., a
file download is faster when allocated more communication resources,
or computation task that can be parallelized, e.g., typical MapReduce jobs \cite{DeG08}, 
and mobile data analytics when additional
compute resources are available \cite{CSK18}. 
For more complex applications involving different types of stages,
the stages conducting massive data processing might be parallelizable, making it possible to 
accelerate by allocating more resources.
For example, in mobile cloud gaming \cite{SRS13}, the most time-consuming 
and resource-consuming stage is usually the cloud rendering
where computing cluster renders the frames of the game. The rendering
procedure can be accelerated by allocating more GPUs, and thus, 
can be viewed as a quantized version of our model.

\begin{figure}[t!]
\includegraphics[width=0.47\textwidth]{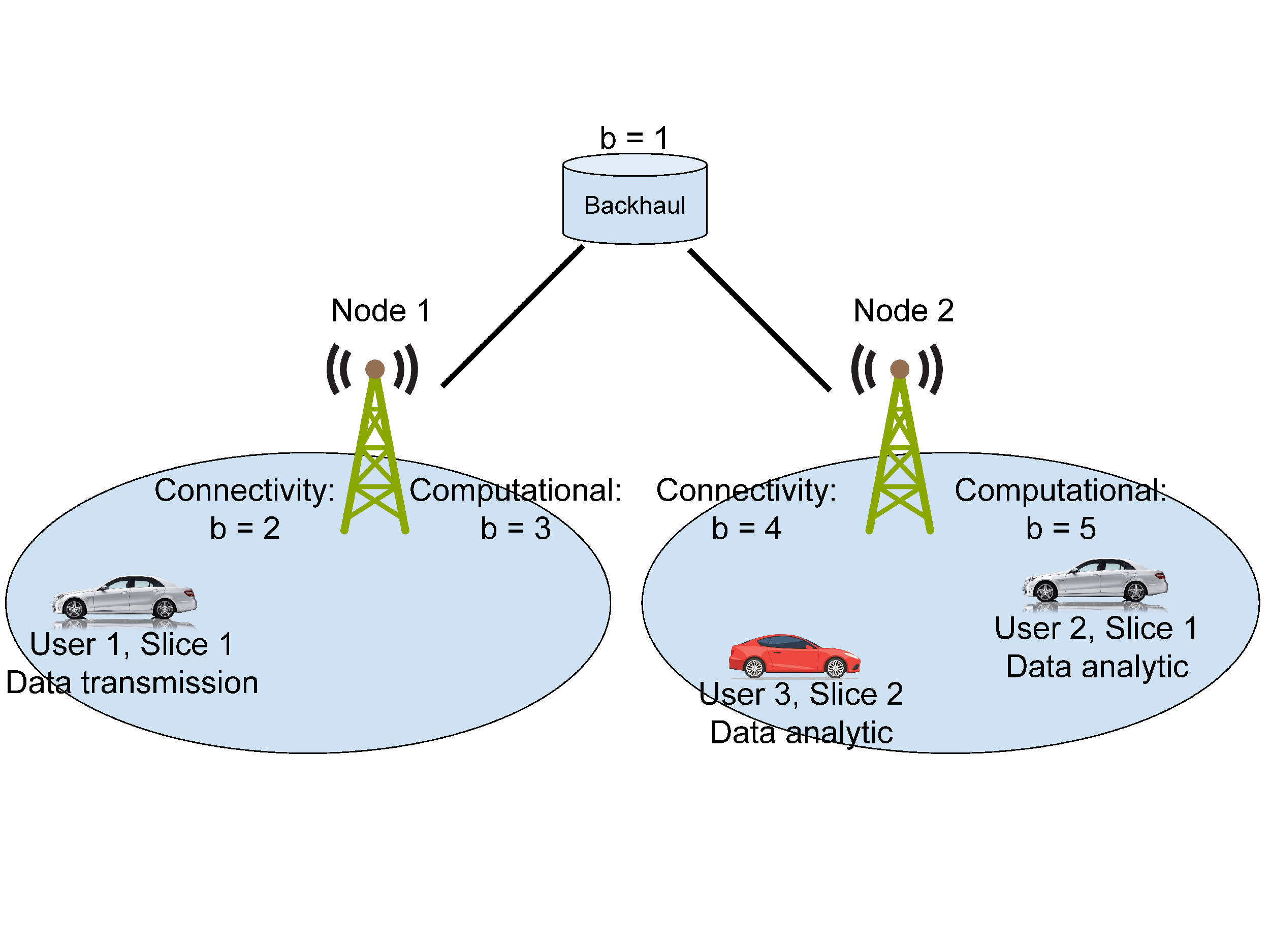}
\caption{Example: network slicing in edge computing with autonomous cars.
}
\label{fig:scf-example}
\end{figure}

\emph{Example: } Let us consider an example where there are two 
autonomous vehicle service operators, say Slice $1$ and 
Slice $2$, coexisting in the same area, and supported by two edge
computing nodes equipped with fronthaul connectivity
and computational resources (e.g., edge GPUs), as shown in Fig.~\ref{fig:scf-example}. 
Both nodes are connected to the same
backhaul node. Different resources at different locations are indexed
as in the figure. There are 3 vehicles (users) in this area,
each of which corresponds to a user class.
Users 1 and 2 are on Slice 1, and User 3 is 
on Slice 2, respectively.
Each autonomous vehicle can run either of two applications.
User 1 is conducting simple data transmission, 
with the resource demand vector 
$\mathbf{d}_1 = (1, 1, 0, 0, 0)$, meaning that
User 1's application involves only connectivity resources,
and to achieve a unit transmission
rate for User 1, the system needs to allocate all the connectivity resources at both 
Node 1 and the backhaul.
Meanwhile, User 2 and 3 are performing mobile data analytics, with demand vectors
$\mathbf{d}_2 = \mathbf{d}_3 = (0.6, 0, 0, 1, 1)$,
meaning that 
to achieve a unit processing rate for Users 2 
or 3, the system needs to allocate 60\%
of the backhaul resource, all the fronthaul resource, together with
all the computational resource at Node 2.
Then, for example, if the 
resource allocation is as given in Table \ref{tab:allocation-example},
the system can achieve user service/processing rates given by
$\lambda_1 = 0.4, \lambda_2 = \lambda_3 = 0.5$.

\begin{table}[t!]
\resizebox{0.48\textwidth}{!}{
\begin{tabular}{|c||l|l|l|l|l|l|}
\hline
\backslashbox{User}{Resource} & 1   & 2   & 3 & 4   & 5   & Rate \\ \hline\hline
user 1 & 0.4 & 0.4 & 0 & 0   & 0   & $\lambda_1 = 0.4$ \\ \hline
user 2 & 0.3 & 0   & 0 & 0.5 & 0.5 & $\lambda_2 = 0.5$ \\ \hline
user 3 & 0.3 & 0   & 0 & 0.5 & 0.5 & $\lambda_3 = 0.5$ \\ \hline
\end{tabular}
}
\caption{Example resource allocation}
\label{tab:allocation-example}
\end{table}

Next, we introduce the concept of \emph{network share}. 
We assign each slice $v$ a positive
share $s_v$ representing the fraction of overall resources to
be committed to
slice $v$. The share allocations across slices are denoted by $\mathbf{s} := (s_v : v\in\mathcal{V})$.
Without loss of generality we assume $\sum_{v\in\mathcal{V}}s_v = 1$.
In turn, each slice distributes its share $s_v$ across its users $u\in\mathcal{U}^v$
according to a \emph{Share-constrained weight allocation} scheme, defined
as follows.
\begin{definition}\label{def:scwa}
{\bf Share-constrained weight allocation (SCWA): }
A weight allocation across users $\mathbf{w} := (w_u : u\in\mathcal{U})$
is a share-constrained weight allocation if for each slice $v$,
\begin{equation}
    \sum_{u\in\mathcal{U}^v} w_u = s_v.
\label{eq:share-constraint}
\end{equation}
\end{definition}
If we consider the weight of each class $c$ 
as $q_c := \sum_{u\in\mathcal{U}_c} w_u$,
Eq. (\ref{eq:share-constraint}) implies $\sum_{c\in\mathcal{C}^v} q_c = s_v$.
As a result, a slice can increase its users' 
weight by purchasing more shares. Also, note that if the number of users 
on a slice surges without increasing the associated share, 
on average each of its users should be given less weight.
Two examples of SCWA are
\begin{enumerate}
\item {\bf equal intra-class weight allocation},
where user weights are the same within a user class, i.e., 
$w_u = \frac{q_c}{n_c}$, for $u\in\mathcal{U}_c$ with
$\sum_{c\in\mathcal{C}^v}q_c = s_v$; and
\item {\bf equal intra-slice weight allocation},
where user weights are the same within a slice, i.e.,
$w_u = \frac{s_v}{n^v}$, for $u\in\mathcal{U}^v$. As a result,
$q_c = \frac{s_{v(c)} n_c}{n^{v(c)}}$. One can see that equal intra-slice 
allocation is a further special case of equal intra-class
allocation. When each user only demands one resource,
such allocation 
emerges naturally as the social optimal, market and Nash equilibrium
when slices exhibit (price taking) strategic behavior
in optimizing their own utility, see \cite{Bra14}.
\end{enumerate}

In turn,
the resources are ultimately committed to users, so a user-level
resource allocation criterion is necessary. 
Let us denote the user rate
allocation by $\bm{\lambda} := (\lambda_u : u\in\mathcal{U})$.
In this paper, we assume equal intra-class weight allocation
is used, resulting in equal rate allocation 
within a user class. Thus a class-level allocation criterion
can be easily converted to a user-level one.
For simplicity, the aggregated rate allocation across user classes is
then denoted by $\bm{\phi} = (\phi_c : c\in\mathcal{C})$,
where $\phi_c := n_c\lambda_u, u\in\mathcal{U}_c$,
and the weight allocation across user classes 
by $\mathbf{q} = (q_c : c\in\mathcal{C})$.
For each slice $v$, the weight allocation (across user classes)
is $\mathbf{q}^v := (q_c : c\in\mathcal{C}^v)$,
and the rate allocation is $\bm{\phi}^v := (\phi_c : c\in\mathcal{C}^v)$.
In view of Eq. (\ref{eq:share-constraint}), we define the normalized weight
allocation for slice $v$ as $\tilde{\mathbf{q}}^v :=
(\tilde{q}_c := \frac{q_c}{s_v} : 
c\in\mathcal{C}^v)$.
The rate allocation across slices is 
$\bm{\gamma} := (\gamma_v := \sum_{c\in\mathcal{C}^v} \phi_c
 : v\in\mathcal{V})$.
The overall rate across
the system is $\lambda := \|\bm{\lambda}\|_1 = \|\bm{\phi}\|_1 = \|\bm{\gamma}\|_1$,
where $\|\cdot\|_1$ is the L1-norm.
SCS is thus defined as follows.
 
\begin{definition}
\textbf{$\alpha-$Share Constrained Slicing ($\alpha-$SCS):}
Under equal intra-class weight allocation with class weights
$\mathbf{q}$,
a class-level rate allocation $\bm{\phi}$ corresponds to $\alpha-$SCS if it is the 
solution to the following problem
\begin{eqnarray}
\max_{\bm{\phi}}~ \{ U_\alpha(\bm{\phi} ; \mathbf{q})
: \sum_{c\in\mathcal{C}_r} d^r_c \phi_c \le 1, \forall r\in\mathcal{R}\},
\label{eq:scf-def}
\end{eqnarray} 
where $\alpha > 0$ is a pre-defined parameter and 
\begin{eqnarray*}
U_\alpha(\bm{\phi} ; \mathbf{q}) := \left\{
\begin{array}{ll}
e^{\sum_{v\in\mathcal{V}}U_\alpha^v(\bm{\phi}^v ;\mathbf{q}^v)} & \alpha = 1\\
\sum_{v\in\mathcal{V}}U_\alpha^v(\bm{\phi}^v ;\mathbf{q}^v) & \alpha > 0~~\mbox{and}~~ \alpha \ne 1,
\end{array}
\right.
\end{eqnarray*}
where $U_\alpha^v(\bm{\phi}^v ;\mathbf{q}^v)$ represents the utility
function of slice $v$ and is given by
\begin{eqnarray*}
U_\alpha^v(\bm{\phi}^v ;\mathbf{q}^v) := \left\{
\begin{array}{ll}
\sum_{c\in\mathcal{C}^v} q_c \log\left(\frac{\phi_c}{q_c}\right) & \alpha = 1\\
\sum_{c\in\mathcal{C}^v} q_c \frac{(\phi_c / q_c)^{1-\alpha}}{1 - \alpha} &
\alpha > 0~~\mbox{and}~~ \alpha \ne 1.
\end{array}
\right.
\end{eqnarray*}
\end{definition}


The criterion underlying SCS
is different from class-level (weighted) $\alpha-$fairness proposed
in \cite{MoW00} and \cite{BoM01}, which is defined as follows.
\begin{definition}
{\bf Class-level $\alpha-$fairness:} Under equal intra-class 
weight allocation, given $\mathbf{q}$, a class-level rate allocation $\bm{\phi}$
corresponds to (weighted) $\alpha-$fairness if it is the solution to 
Problem (\ref{eq:scf-def}) with utility function of slice $v$ 
given by 
\begin{eqnarray*}
U_\alpha^v(\bm{\phi}^v ;\mathbf{q}^v) := \left\{
\begin{array}{ll}
\sum_{c\in\mathcal{C}^v} q_c \log\left(\phi_c\right) & \alpha = 1\\
\sum_{c\in\mathcal{C}^v} q_c \frac{(\phi_c)^{1-\alpha}}{1 - \alpha} &
\alpha > 0~~\mbox{and}~~ \alpha \ne 1.
\end{array}
\right.
\end{eqnarray*}
\end{definition}
As shown in \cite{MoW00}, $\alpha-$fairness is equivalent to 
(weighted) proportional fairness as $\alpha = 1$
and unweighted maxmin fairness as $\alpha \rightarrow \infty$,
while the asymptotic characterization of $\alpha-$SCS is given as follows.
\begin{corollary}
$\alpha-$SCS
is equivalent to (weighted) proportional fairness as $\alpha = 1$, 
and \emph{weighted} max-min fairness as $\alpha \rightarrow \infty$.

Here under equal intra-class weight allocation, 
weighted proportional fairness is defined as the solution to 
the following problem:
\begin{eqnarray}
\max_{\bm{\phi}}~\left\{
\sum_{c\in\mathcal{C}} q_c \log \phi_c :
\sum_{c\in\mathcal{C}_r} d^r_c \phi_c \le 1, \forall r\in\mathcal{R}
\right\},
\end{eqnarray}
and weighted max-min fairness is defined as the solution to the 
following problem:
\begin{eqnarray}
\max_{\bm{\phi}} ~\left\{
\min_{c\in\mathcal{C}} \frac{\phi_c}{q_c} :
\sum_{c\in\mathcal{C}_r} d^r_c \phi_c \le 1, \forall r\in\mathcal{R}
\right\}.
\end{eqnarray}
\end{corollary}

The persistence of weight is 
important, especially when $\alpha$ increases. 
Otherwise, the notion of share does not matter
when $\alpha$ is large, undermining inter-slice protection.
To the best of our knowledge, SCS is the first variation of $\alpha-$fairness 
incorporating user weighting in a consistent manner.

\iffull
\begin{IEEEproof}
When $\alpha = 1$, one can see that the maximum is assumed when 
$\sum_{c\in\mathcal{C}} q_c \log \left(\frac{\phi_c}{q_c}\right)$ assumes maximum.
Due to the concavity, a rate allocation $\bm{\phi}^* := (\phi_c^* : c\in\mathcal{C})$ is the maximizer
if and only if 
\begin{eqnarray*}
\sum_{c\in\mathcal{C}} \frac{q_c}{\phi_c^*} (\phi_c^\prime - \phi_c^*) \le 0, 
\end{eqnarray*}
for any feasible $\bm{\phi}^\prime$.
Also, for $\alpha-$SCS with weight $\mathbf{q}$, 
when $\alpha \ne 1$, $\bm{\phi}^*$ is the maximizer
if and only if 
\begin{eqnarray*}
\sum_{c\in\mathcal{C}} \left(\frac{\phi_c^*}{q_c}\right)^{-\alpha} (\phi_c^\prime - \phi_c^*) \le 0, 
\end{eqnarray*}
for any feasible $\bm{\phi}^\prime$. One can see that two optimality
conditions coincide when $\alpha = 1$.

The asymptotic behavior when $\alpha \rightarrow \infty$
is a direct corollary of the Lemma 3 in \cite{MoW00}.
\end{IEEEproof}
\else
\fi

Let us define function $f_\alpha(\mathbf{x} ; \mathbf{y})$ of two positive 
vectors $\mathbf{x},\mathbf{y} \in \mathbb{R}^n_{+}$ 
such that $\|\mathbf{x}\|_1 = \|\mathbf{y}\|_1 = 1$ as
\begin{eqnarray}
f_\alpha(\mathbf{x} ; \mathbf{y}) = \left\{
\begin{array}{ll}
e^{-D_{KL}(\mathbf{x} \| \mathbf{y})} & \alpha = 1 \\
\left(\sum_{i} x_i \left(\frac{y_i}{x_i}\right)^{1-\alpha}\right)^{\frac{1}{\alpha}} & \alpha > 0, \alpha \ne 1,
\end{array}
\right.
\end{eqnarray}
where $D_{KL}(\cdot \| \cdot)$ represents the Kulback-Leibler (K-L) divergence.
The function $f_\alpha(\mathbf{x};\mathbf{y})$
can be viewed as a measure of how close a normalized 
resource allocation $\mathbf{x}$ is to a normalized weight vector $\mathbf{y}$
in that, for example, 
when $\alpha = 1$, it decreases with the K-L divergence between $\mathbf{x}$
and $\mathbf{y}$, thus assumes maximum when $\mathbf{x} = \mathbf{y}$,
meaning that the rate allocation is aligned with the specified weights.
Thus $f_\alpha(\mathbf{x};\mathbf{y})$ can
be interpreted as a measure of $\mathbf{y}-$weighted fairness of allocation $\mathbf{x}$.

\iffull
One can easily show that $f_\alpha(\mathbf{x} ; \mathbf{y})$ is continuous.
Moreover, for general $\alpha$ and a given $\mathbf{y}$, 
the following claim can be shown by setting
the partial derivative of the associated Lagrangian to 0.
\begin{proposition}
$f_\alpha(\mathbf{x} ; \mathbf{y})$ assumes maximum when the rate is aligned with the weight
    	when no constraint is imposed, i.e.,
    	\begin{equation}
			f_\alpha(\mathbf{y}; \mathbf{y}) = \max_{\mathbf{x}} f_\alpha(\mathbf{x};\mathbf{y})
		\end{equation}
\end{proposition}
%
\else
\fi

One can show that for a given $\alpha$,
$\alpha-$SCS criterion
can be factorized as follows.
\begin{proposition}
For the $\alpha-$SCS criterion,
\begin{equation}
U_\alpha(\bm{\phi} ; \mathbf{q})
= E_\alpha(\lambda) \left(F^{\textrm{inter}}_\alpha(\bm{\gamma}) 
F^{\textrm{intra}}_\alpha(\bm{\phi} ; \mathbf{q})\right)^\alpha,
\label{eq:scf-factorization}
\end{equation}
where $E_\alpha(\lambda)$, $F^{\textrm{inter}}_\alpha(\bm{\gamma})$
and $F^{\textrm{intra}}_\alpha(\bm{\phi} ; \mathbf{q})$ can be
interpreted as the overall network efficiency, inter-slice 
and intra-slice fairness, respectively.
\end{proposition}

In Eq. (\ref{eq:scf-factorization}), the efficiency is captured by 
a concave non-decreasing function of $\lambda$ given by
\begin{eqnarray*}
E_\alpha(\lambda) := \left\{
\begin{array}{ll}
\lambda & \alpha = 1\\
\frac{\lambda^{1 - \alpha}}{1 - \alpha} & \alpha > 0~~\mbox{and}~~ \alpha \ne 1.
\end{array}
\right.
\end{eqnarray*}
The inter-slice fairness function is given by
\begin{eqnarray*}
F^{\textrm{inter}}_\alpha(\bm{\gamma})
:= f_\alpha(\tilde{\bm{\gamma}} ; \mathbf{s}),
\end{eqnarray*}
where $\tilde{\bm{\gamma}}: = (\tilde{\gamma}^v:= {\gamma^v}/{\lambda} : v\in\mathcal{V})$ is
the normalized aggregated rate across slices.
Let us define the normalized rate allocation across user classes 
on slice $v$ as $\tilde{\bm{\phi}}^v := (\tilde{\phi}_c:=\frac{\phi_c}{\gamma^v}
 : c\in\mathcal{C}^v)$. The intra-slice fairness term is then given by
\begin{eqnarray*}
F^{\textrm{intra}}_\alpha(\bm{\phi} ; \mathbf{q})
:= \left\{
\begin{array}{ll}
e^{\sum_{v\in\mathcal{V}} t^v_\alpha(\tilde{\bm\gamma} ; \mathbf{s}) 
\log f_\alpha(\tilde{\bm{\phi}}^v ; \tilde{\mathbf{q}}^v)} & \alpha = 1 \\
\left(\sum_{v\in\mathcal{V}} t^v_\alpha(\tilde{\bm\gamma} ; \mathbf{s}) 
(f_\alpha(\tilde{\bm{\phi}}^v ; \tilde{\mathbf{q}}^v))^\alpha
\right)^{\frac{1}{\alpha}} & \alpha \ne 1,
\end{array}
\right.
\end{eqnarray*}
where $t^v_\alpha(\tilde{\bm\gamma} ; \mathbf{s})$ can be viewed as 
the weight for the fairness of each slice $v$:
\begin{equation}
t^v_\alpha(\tilde{\bm\gamma} ; \mathbf{s}) 
:= \frac{s_v (\frac{\tilde{\gamma}^v}{s_v})^{1 - \alpha}}
{\sum_{v^\prime \in\mathcal{V}} s_{v^\prime} (\frac{\tilde{\gamma}^{v^\prime}}{s_{v^\prime}})^{1-\alpha}}.
\end{equation}

One can see that Eq. (\ref{eq:scf-factorization}) captures a
trade-off among overall network efficiency, inter-slice fairness, 
which can be seen as a proxy of inter-slice protection, 
and intra-slice fairness. The significance of fairness increases
as $\alpha$ increases. When $\alpha \rightarrow 0$, $\alpha-$SCS
is maximizing the overall rate allocated, regardless
of the weights.
In order to achieve desirable resource utilization,
a sharing criterion should realize \emph{load-driven elasticity},
i.e., the amount of resources provisioned to a user class
increases in the number of its users.
Under equal intra-slice weight allocation, 
from Eq. (\ref{eq:scf-factorization}) one can observe that,
due to the fairness terms, the relative resource allocation
of a slice tends to be aligned with $\tilde{\mathbf{q}}^v
= (\frac{n_c}{n^v} : c\in\mathcal{C}^v)$, i.e., its relative
load distribution. Thus the elasticity of $\alpha-$SCS is achieved
as a result of weighted fairness.
Specifically
under SCS and parallel resource assumption,
i.e., each user only uses one resource, $|\mathcal{R}_c| = 1,\forall c\in\mathcal{C}$,
one can show the following result.

\begin{theorem}\label{thm:elasticity}
Under equal intra-slice weight allocation, assuming
$|\mathcal{R}_c| = 1,\forall c\in\mathcal{C}$,
$\alpha-$SCS is such that $\phi_c$ is a monotonically increasing
function of $n_c$, when $n_{c^\prime}$ is fixed for $c^\prime \ne c$.
\end{theorem}

Specifically in the setting of \Cref{thm:elasticity},
each resource $r$ will provision its resource
across user classes in proportion to $\frac{s_{v(c)}n_c}{n^{v(c)}}$.


Such elasticity is key to achieving a sharing scheme that is 
aware of the inter-slice protection, while still improves the 
resource utilization by accommodating dynamic user loads on different
slices.


\iffull
\section{Static Analysis}
\label{sec:static}
\subsection{System model}
In this section we will take a closer look at the characterization 
of SCS slice level rate allocations.

The SCS criterion (Problem (\ref{eq:scf-def})) is 
equivalent to the solution to the following problem
\begin{eqnarray}
\max_{\bm{\phi}} ~\left\{
\sum_{v\in\mathcal{V}} U_\alpha^v(\bm{\phi}^v ; \mathbf{q}^v) : 
\sum_{c\in\mathcal{C}_r} d^r_c \phi_c \le 1,~~\forall r\in\mathcal{R}
\right\}.
\label{prob:rate-inelastic}
\end{eqnarray}

We shall explore two key desirable properties for a
sharing criterion, namely, \emph{protection} and \emph{envyfreeness}.
In our setting, protection means that no slice is penalized under 
SCS sharing vs. static partitioning.
Envyfreeness means that no slice is motivated to swap its resource
allocation with another slice with a smaller share. 
These two properties together motivate the choice of $\alpha-$SCS sharing,
and at least partially purchasing a larger share
in order to improve performance.

%

\subsection{Protection}
Formally, let us characterize protection among slices by how much
performance deterioration is possible for a slice when switching
from static partitioning to $\alpha-$SCS sharing. Note that under static partitioning,
slices are decoupled, so inter-slice protection is achieved possibly
at the cost of efficiency. To be specific, the rate allocation
for slice $v$ under static partitioning is given by the
following problem.
\begin{eqnarray}
\max_{\bm{\phi}^v} ~\left\{
U_\alpha^v(\bm{\phi}^v ; \mathbf{q}^v):
 \sum_{c\in\mathcal{C}^v_r}d^r_c\phi_c \le s_v,~~\forall r\in\mathcal{R}\right\},
 \label{eq:static-partition}
\end{eqnarray}

From now on, for a given $\alpha$,
let us denote the rate allocation for slice
$v$ under $\alpha-$SCS by $\bm{\phi}^{v,S} := (\phi^S_c : c\in\mathcal{C}^v)$,
and that under static
partitioning by $\bm{\phi}^{v,P} := (\phi^P_c : c\in\mathcal{C}^v)$.
The parameter $\alpha$ is suppressed when there is no ambiguity. 
The following result
demonstrates that $\alpha-$SCS with $\alpha = 1$ achieves inter-slice 
protection in that any slice achieves a better utility
under $\alpha-$SCS sharing.

\begin{theorem}\label{thm:protection}
For a given $\mathbf{q}$, 
when the resource allocation is performed according
to $\alpha-$SCS,
difference in slice $v$'s utility compared to that under
static partitioning
is upper-bounded by (when $\alpha \ne 1$)
\begin{eqnarray*}
\lefteqn{U^\alpha_v(\bm{\phi}^{v, P}; \mathbf{q}^v) - U^\alpha_v(\bm{\phi}^{v, S}; \mathbf{q}^v)
\le } \nonumber\\
& & s_v\left( \sum_{c\in\mathcal{C}^v}\tilde{q}_c (\sum_{r\in\mathcal{R}_c} d^r_c\nu^*_r )^{\frac{\alpha - 1}{\alpha}}  
 -\sum_{c\in\mathcal{C}}q_c (\sum_{r\in\mathcal{R}_c} d^r_c\nu^*_r )^{\frac{\alpha-1}{\alpha}}
 \right),
\end{eqnarray*}
where $\tilde{q}_c := q_c / s_{v(c)}$ is the
normalized weight of class $c$,
and $\nu^*_r$ is the optimal dual variable associated with
the capacity constraint at resource $r$ in Problem (\ref{prob:rate-inelastic}), also known as
the shadow price of resource $r$.
\end{theorem}

\textbf{Remark:} The right hand side characterizes how the protection
changes with $\alpha$.
$\sum_{c\in\mathcal{C}^v}\tilde{q}_c (\sum_{r\in\mathcal{R}_c} d^r_c\nu^*_r )^{1 - \frac{1}{\alpha}}$
can be viewed as the average of the $(1-\frac{1}{\alpha})-$order moment of `charged' resource
usage of slice $v$'s user,
while $\sum_{c\in\mathcal{C}}q_c (\sum_{r\in\mathcal{R}_c} d^r_c\nu^*_r )^{1 - \frac{1}{\alpha}}$
is that of the overall users.
When $0 < \alpha < 1$, sharing tends to benefit slices with greater
average user usages, at the cost
of other slices, while when $\alpha > 1$, slices with smaller average
user prices are preferred.

When $\alpha \rightarrow 1$, the utility of slice $v$,
$U_v^\alpha(\bm{\phi}^v ; \mathbf{q}^v)
= \frac{1}{1-\alpha} \\ \times\sum_{c\in\mathcal{C}^v} q_c 
\left(\frac{\phi_c}{q_c}\right)^{1-\alpha}$
tends to be non-changing with $\bm{\phi}$, so the result in \Cref{thm:protection}
seems to be trivial. However, due to the factor $\frac{1}{1-\alpha}$,
the utility function is not well-defined at $\alpha \rightarrow 1$. Thus the exact
result for $\alpha$ need to be discussed on its own, as in \Cref{thm:protection-log}.

Also, note that 
when SCWA constraint Eq. (\ref{eq:share-constraint}) is voided,
another form of the theorem can be written as 
$$
U^\alpha_v(\bm{\phi}^{v, P}; \mathbf{q}^v) 
- U^\alpha_v(\bm{\phi}^{v, S}; \mathbf{q}^v)
\le w_v (p_v - s_v \frac{w_0}{w_v} \bar{p}_0),
$$
where 
$w_v := \sum_{c\in\mathcal{C}^v} q_c$ is the total weight of users of slice $v$,
$w_0 := \sum_{c\in\mathcal{C}} q_c$ is the total weight of all users, 
$p_v := \sum_{c\in\mathcal{C}^v} \tilde{q}_c \left(\sum_{r\in\mathcal{R}_c}d^r_c\nu^*_r\right)^{1-\frac{1}{\alpha}}$
is the average $1-\frac{1}{\alpha}$ power of the prices of slice $v$'s users, and
$\bar{p}_0 := \sum_{v} \frac{w_v}{w_0} p_v$ is the weighted average of $p_v$ across
all slices. One could see that,
if $w_v$ is not limited, as $w_v \rightarrow \infty$, as long as 
$p_v \ne \bar{p}_0$, the gap can be arbitrarily bad, implying
significant utility loss when slice-level sharing is used. In comparison,
If we use SCWA,
i.e., constrained by Eq.~(\ref{eq:share-constraint}),
the right hand side equals to $s_v (p_v - \bar{p}_0)$.
This quantity is small when $s_v$ is small,
or slice $v$'s users do not use many `expensive'
resources.

\iffull
\begin{IEEEproof}
Under sharing scheme, the rate allocation for each slice $v$ should be the
same as the solution to the following problem:
\begin{eqnarray}
\max_{\bm{\phi}^v} &&
U_\alpha^v(\bm{\phi}^v ; \mathbf{q}^v) \nonumber\\
\textrm{such that } && \sum_{c\in\mathcal{C}^v_r}d^r_c\phi_c \le \sum_{c\in\mathcal{C}^v_r}d^r_c\phi_c^{S},~~\forall r\in\mathcal{R}.
\label{problem:slice-sharing}
\end{eqnarray}
Problem (\ref{problem:slice-sharing}) yields solution $\bm{\phi}^{v, \textrm{S}}$.
In comparison, under static partitioning scheme,
the rate allocation for slice $v$ is given by
\begin{eqnarray}
\max_{\bm{\phi}^v} &&
U_\alpha^v(\bm{\phi}^v ; \mathbf{q}^v) \nonumber\\
\textrm{such that } && \sum_{c\in\mathcal{C}^v_r}d^r_c\phi_c \le s_v,~~\forall r\in\mathcal{R},
\label{problem:slice-partitioning}
\end{eqnarray}
which can be regarded as a perturbed version of Problem (\ref{problem:slice-sharing}),
and yields solution $\bm{\phi}^{v, \textrm{P}}$.
It is a well known result in convex optimization \cite{BoV04} that the
change in the optimal objective function value
due to perturbation of the constraints
can be bounded by:
\begin{equation}
U^\alpha_v(\bm{\phi}^{v, \textrm{P}}; \mathbf{q}^v) - U^\alpha_v(\bm{\phi}^{v, \textrm{S}}; \mathbf{q}^v)
\le \sum_{r\in\mathcal{R}} \nu^*_r \left( \sum_{c\in\mathcal{C}^v_r}d^r_c\phi_c^{S} - s_v \right).
\label{eq:cvx-sensitivity}
\end{equation}

The Lagrangian of Problem (\ref{problem:slice-sharing}) is
\begin{eqnarray}
\lefteqn{L^\alpha(\bm{\phi}^v; \boldsymbol{\nu})
= -\frac{1}{1-\alpha}\sum_{c\in\mathcal{C}^v} q_c \left( \frac{\phi_c}{q_c} \right)^{1-\alpha}}\nonumber\\
& & + \sum_{r\in\mathcal{R}}
\nu_r (\sum_{c\in\mathcal{C}^v_r}d^r_c\phi_c - \sum_{c\in\mathcal{C}^v_r}d^r_c\phi_c^{S}).
\end{eqnarray}
Setting the partial derivative against $\bm{\phi}^v$ to 0, we obtain the dual function as:
\begin{eqnarray*}
\lefteqn{g^\alpha(\boldsymbol{\nu})
= -\frac{1}{1 - \alpha} \sum_{c\in\mathcal{C}^v} q_c \left( \sum_{r\in\mathcal{R}_c}d^r_c\nu_r \right)^{1- \frac{1}{\alpha}}} \nonumber\\
& & + \sum_{r\in\mathcal{R}} \nu_r \left( \sum_{c\in\mathcal{C}^v_r} d^r_c q_c \left(\sum_{r^\prime\in\mathcal{R}_c}d^{r^\prime}_c\nu_{r^\prime}\right)^{-\frac{1}{\alpha}}
- \sum_{c\in\mathcal{C}^v_r} d^r_c\phi_c^S
 \right).
\end{eqnarray*}
Also,
\begin{equation}
\phi_c^S =  q_c \left(\sum_{r^\prime\in\mathcal{R}_c}d^{r^\prime}_c\nu^*_{r^\prime}\right)^{-\frac{1}{\alpha}}.
\label{eq:rus-as-optimal-dual}
\end{equation}
By swapping the order of summation, one can show that
\begin{eqnarray*}
\lefteqn{\sum_{r\in\mathcal{R}} \nu_r \sum_{c\in\mathcal{C}^v_r} d^r_c q_c \left(\sum_{r^\prime\in\mathcal{R}_c}d^{r^\prime}_c\nu_{r^\prime}\right)^{-\frac{1}{\alpha}}
=} \\
& & \sum_{c \in \mathcal{C}^v} q_c \left(\sum_{r\in\mathcal{R}_c} d^r_c \nu_r \right)^{1 - \frac{1}{\alpha}}.
\end{eqnarray*}

Due to strong convexity of Problem (\ref{problem:slice-sharing}), we know
$$
g^\alpha(\bm{\nu}^*) = -\frac{1}{1-\alpha} \sum_{c\in\mathcal{C}^v}
q_c \left( \frac{\phi_c^S}{q_c}\right)^{1-\alpha}.
$$

Plugging in Eq. (\ref{eq:rus-as-optimal-dual}) we have
\begin{equation}
\sum_{c \in \mathcal{C}^v}q_c \left(\sum_{r\in\mathcal{R}_c} d^r_c \nu_r^* \right)^{1 - \frac{1}{\alpha}}
- \sum_{r\in\mathcal{R}}\nu^*_r \sum_{c\in\mathcal{C}^v_r} d^r_c\phi_c^S = 0.
\label{eq:rate-and-shadow-price}
\end{equation}
Note that if
a resource is binding, the sum of resource allocated should equal to 1. Otherwise it
has 0 shadow price.
Summing above across $v\in\mathcal{V}$, we have
\begin{equation}
\sum_{c\in\mathcal{C}} q_c \left(\sum_{r\in\mathcal{R}_c} d^r_c\nu_r^* \right)^{1 - \frac{1}{\alpha}}
- \sum_{r\in\mathcal{R}} \nu_r^* = 0.
\end{equation}
Plugging in the right hand side of Eq. (\ref{eq:cvx-sensitivity}) 
to substitute $\sum_{r\in\mathcal{R}} \nu_r^*$, and also plugging in 
Eq. (\ref{eq:rus-as-optimal-dual}),
the theorem is proved.
\end{IEEEproof}
\else
\fi

Following is the result specifically for the case when $\alpha = 1$.
\begin{theorem}\label{thm:protection-log}
For a given $\mathbf{q}$, when the resource allocation
is performed according to $1-$SCS,
slice $v$'s utility exceeds that under
static partitioning (Problem (\ref{eq:static-partition})),
i.e., 
\begin{equation}
U^v_1(\bm{\phi}^{v, P}; \mathbf{q}^v) \le U^v_1(\bm{\phi}^{v, S}; \mathbf{q}^v).
\end{equation}
\end{theorem}

\textbf{Remark:} It is a straightforward observation that
under $\alpha-$SCS, the global utility 
$\sum_{v\in\mathcal{V}} U^v_1(\bm{\phi}^v ; \mathbf{q}^v)$
is improved since it can be viewed as relaxing the system
constraints. However, Theorem \ref{thm:protection-log} asserts
that this holds uniformly on a per slice basis.

\iffull
\begin{IEEEproof}
Similar to the argument for general $\alpha$, we have that the gap between
sharing and static partitioning satisfies Eq. (\ref{eq:cvx-sensitivity}).
Also, by solving the first order condition, one can obtain that
$\phi^S_c = \frac{q_c}{\sum_{r\in\mathcal{R}_c} d^r_c \nu_r^*}$.
By plugging in this expression and swapping the order of summation we have
\begin{equation}
U^v_1(\bm{\phi}^{v, P}; \mathbf{q}^v) - U^v_1(\bm{\phi}^{v, S}; \mathbf{q}^v)
\le s_v(1 - \sum_{r\in\mathcal{R}}\nu_r^*),
\label{eq:log-protection-intermediate}
\end{equation}
where $\nu_r^*$ is the shadow price of resource $r$
under SCS, or the dual variables
associated with the capacity constraints.

Then if we have $\sum_{r\in\mathcal{R}}\nu_r^* = 1$, the proof is complete.
For $\alpha = 1$, the Lagrangian is given by
$$
L^1(\bm{\phi}^v; \boldsymbol{\nu}) = -\sum_{c\in\mathcal{C}^v} q_c \log \phi_c + \sum_{r\in\mathcal{R}}
\nu_r (\sum_{c\in\mathcal{C}^v_r}d^r_c\phi_c - \sum_{c\in\mathcal{C}^v_r}d^r_c\phi_c^{S}).
$$
By setting the derivative against $\bm{\phi}^v$ to 0, we have the dual function as
$$
g^1(\boldsymbol{\nu}) = -\sum_{c\in\mathcal{C}^v} q_c \log\frac{q_c}{\sum_{r\in\mathcal{R}_c}d^r_c\nu_r} 
+ s_v - \sum_{r\in\mathcal{R}} \nu_r \sum_{c\in\mathcal{C}^v_r}d^r_c\phi_c^{S},
$$
and
\begin{equation}
\phi_c^S = \frac{q_c}{\sum_{r\in\mathcal{R}_c} d^r_c\nu^*_r}.
\label{eq:log-ru}
\end{equation}
By strong duality, maximal dual should be minimal primal function. And optimal dual is achieved
at the shadow price $\boldsymbol{\nu}^*$. Thus,
$$
g^1(\boldsymbol{\nu}^*) =  -\sum_{c\in\mathcal{C}^v} q_c \log \phi_c^{S},
$$
which gives us
\begin{equation}
s_v - \sum_{r\in\mathcal{R}}\nu^*_r\sum_{c\in\mathcal{C}^v_r}d^r_c\phi_c^{S} = 0.
\label{eq:log-intermediate-result}\end{equation}
Summing above across $v\in\mathcal{V}$ we have $1 - \sum_{r\in\mathcal{R}}\nu^*_r = 0$. Because if
a resource is binding, the sum of rate allocated should be equal to 1. Otherwise it
has 0 shadow price. 
Plugging above result into Eq. (\ref{eq:log-protection-intermediate}),
the theorem is proved. 
\end{IEEEproof}
\else
\fi

\subsection{Envyfreeness}
Formally, envyfreeness is defined under the assumption that,
for two slices $v$ and $v^\prime$,
if they swap their allocated resources, slice $v$'s associated utility
will not be improved if $s_{v^\prime} \le s_v$.
Before swapping, the rate allocation for slice $v$ is given by
$\bm{\phi}^{v, S}$, while after
swapping with slice $v^\prime$, its rate allocation is determined by
solving following problem:
\begin{eqnarray*}
\max_{\bm{\phi}^v}~\{U^v_\alpha(\bm{\phi}^v; \mathbf{q}^v) : 
\sum_{c\in\mathcal{C}^v_r} d^r_c \phi_c \le 
\sum_{c\in\mathcal{C}^{v^\prime}_r} d^r_c \phi_c^S,
~\forall r\in\mathcal{R}\}.
\end{eqnarray*}

Note that $\sum_{c\in\mathcal{C}^{v^\prime}_r} d^r_c \phi_c^S$
corresponds to the fraction of resource $r$ provisioned to slice $v^\prime$.
Let us denote the solution to such problem for slice $v$
as $\bm{\phi}^{v\leftrightarrow v^\prime}$.
Then we have following result.

\begin{theorem}
The difference between the utility obtained by slice $v$
under $\alpha-$SCS with SCWA, and that
under static partitioning within the resource provisioned to 
another slice $v^\prime$
is upper-bounded by the following inequality:
\begin{eqnarray*}
\lefteqn{U^v_\alpha(\bm{\phi}^{v\leftrightarrow v^\prime}; \mathbf{q}^v) -
U^v_\alpha(\bm{\phi}^{v, S}; \mathbf{q}^v) \le} \nonumber\\
& & \sum_{c\in\mathcal{C}^{v^\prime}} q_c \left(\sum_{r\in\mathcal{R}_c}d^r_c\nu^*_r\right)^{\frac{\alpha-1}{\alpha}}
- \sum_{c\in\mathcal{C}^{v}} q_c \left(\sum_{r\in\mathcal{R}_c}d^r_c\nu^*_r\right)^{\frac{\alpha-1}{\alpha}}.
\end{eqnarray*}
\end{theorem}

{\bf Remark:} 
As a special case, when $\alpha = 1$, the right hand side of the inequality 
becomes $s_{v^\prime} - s_v$, and thus a slice has no incentive
to swap its allocation with another
with a less or equal share,
which implies SCS is envyfree.
Envyfreeness implies that $\alpha-$SCS achieves desirable resource 
utilization in that the right portion of resource is provisioned
to the right slice.

\iffull
\begin{IEEEproof}
Still by the sensitivity of convex optimization problem \cite{BoV04}, we have
\begin{eqnarray}
\lefteqn{U^v_\alpha(\bm{\lambda}^{v\leftrightarrow v^\prime} ; \mathbf{q}^v)
- U^v_\alpha(\bm{\lambda}^{v,S} ; \mathbf{q}^v) \le} \nonumber\\
& & \sum_{r\in\mathcal{R}} \nu^*_r \left( \sum_{c\in\mathcal{C}^{v^\prime}_r} d^r_c \phi^S_c
- \sum_{c\in\mathcal{C}^{v}_r} d^r_c \phi^S_c
\right).
\end{eqnarray}

Then by substituting Eq. (\ref{eq:rate-and-shadow-price}) the theorem is proved.
\end{IEEEproof}
\else
\fi

\subsection{Using $\infty-$SCS as a surrogate for $1-$SCS}
From previous discussions, one can see that it is of particular interest to use
$1-$SCS as the fairness criterion, for it achieves strict protection and 
envyfreeness. When $\alpha = 1$, $\alpha-$SCS becomes weighted proportional fairness,
whose solution usually involves iterative methods, and the complexity 
increases rapidly with the number of user classes as well as the accuracy 
requirement, see, e.g., \cite{PiM04}, making it hard to implement in 
large-scale.
In comparison, weighted max-min is relatively easy to implement in distributed manner,
see \cite{GZH11} for example. Specifically a progressive water-filling algorithm
\cite{PiM04} has $O(|\mathcal{C}|\max_{c\in\mathcal{C}}|\mathcal{R}_c|)$ complexity.
Thus, in our work we will discuss the feasibility of using $\infty-$SCS,
which is equivalent to a (dynamically) weighted maxmin,
as a surrogate to $1-$SCS.
If the resulted utility function is not far from the 
optimum of $1-$SCS criterion,
we shall assert $\infty-$SCS achieves similar performance as
$1-$SCS.

For simplicity, we consider the original
form of weighted-log utility, given by
\begin{equation}
\Psi(\bm{\phi} ; \mathbf{q}) := \sum_{c\in\mathcal{C}} q_c \log \phi_c.
\end{equation} 
Then for the overall utility achieved, we have following theorem.
\begin{theorem}
For a given weight allocation $\mathbf{q}$, if $d^r_c\ge 1, \forall r\in\mathcal{R},c\in\mathcal{C}$, 
we have
\begin{equation}
\Psi(\bm{\phi}^{*,1} ; \mathbf{q}) - \Psi(\bm{\phi}^{*,\infty};\mathbf{q})
\le \sum_{c\in\mathcal{C}} q_c D_c - 1, \label{eq:sharing-transferring}
\end{equation}
where $\bm{\phi}^{*,\alpha} := (\phi^{*,\alpha}_c: c\in\mathcal{C})$ is the optimal rate allocation under $\alpha-$SCS, 
and $D_c := \sum_{r\in\mathcal{R}_c} d^r_c$.
\end{theorem}

\emph{Remark:} First note that the condition $d^r_c \ge 1$
can be easily satisfied by rescaling the unit of rate without
loss of generality.
Also by rescaling, one can show that 
such bound vanishes when each user class is associated
with only one resource, i.e., $|\mathcal{R}_c| = 1, \forall c\in\mathcal{C}$,
and $d^r_c$ are the same, e.g., $d^r_c = 1$.
Such bound implies that, the suboptimality due to using
a surrogate solution to achieve weighted proportional fairness
depends on the diversity in the users' requirements on resources.
Also, this gap of suboptimality cannot be arbitrarily bad 
because under SCWA,
we have $\sum_c q_c = 1$, thus the right hand side of 
Eq. (\ref{eq:sharing-transferring}) is at most
$\max_{c} D_c - 1$.

\iffull
\begin{IEEEproof}
Note that when $\alpha \rightarrow \infty$,
SCS approaches weighted maxmin, which can be solved
by a progressive water-filling algorithm.
Let us denote the
resource where class $c$ is bottlenecked under weighted maxmin
by $r(c)$, and in turn, the set of users being bottlenecked at
resource $r$ by $\tilde{\mathcal{C}}_r$
. Let us define $\nu^*_r$ as the shadow price for resource $r$ when $\alpha = 1$.
According to the definition we have
\begin{eqnarray*}
\lefteqn{\Psi(\bm{\phi}^{*,1} ; \mathbf{q}) - \Psi(\bm{\phi}^{*,\infty};\mathbf{q})} \\
& =  & \sum_c q_c (\log \phi_c^{*,1} - \log \phi_c^{*,\infty}) \\
 & \le & \sum_c q_c \left( \log \frac{q_c}{\sum_{r^\prime\in\mathcal{R}_c}d_c^{r^\prime}\nu_{r^\prime}^*} \right.\\
& & - \left.\log \left( \frac{q_c}{\sum_{c^\prime \in \mathcal{C}_{r(c)}}d_{c^\prime}^{r(c)}q_{c^\prime}} \right)\right) \\
 & = & \sum_{r\in\mathcal{R}} \sum_{c\in\tilde{\mathcal{C}}_r} q_c
 \log\left( \frac{\sum_{c^\prime \in \mathcal{C}_{r}}d_{c^\prime}^r q_{c^\prime}}
 {\sum_{r^\prime\in\mathcal{R}_c}d_c^{r^\prime}\nu_{r^\prime}^*} \right).
\end{eqnarray*}
The first inequality follows from the form of solution of
sharing problem when $\alpha = 1$, and the fact that
$\phi_c^{*,\infty} \ge \frac{q_c}{\sum_{c^\prime \in \mathcal{C}_{r(c)}}d_{c^\prime}^{r(c)}q_{c^\prime}}$,
since the worst rate user $u$ could obtain is when there is no other users get saturated
before it at its bottleneck resource. Because $\log x \le x - 1$ we have
\begin{eqnarray*}
\lefteqn{\Psi(\bm{\phi}^{*,1} ; \mathbf{q}) - \Psi(\bm{\lambda}^{*,\infty};\mathbf{q})} \\
&\le & \sum_{r\in\mathcal{R}} \sum_{c\in\tilde{\mathcal{C}}_r} q_c
\frac{\sum_{c^\prime \in \mathcal{C}_{r}}d_{c^\prime}^r q_{c^\prime}}
{\sum_{r^\prime\in\mathcal{R}_c}d_c^{r^\prime}\nu_{r^\prime}^*}  - 1\\
& = & \sum_{r\in\mathcal{R}} \left(\sum_{c^\prime \in \mathcal{C}_{r}}d_{c^\prime}^r q_{c^\prime}\right)
 \sum_{c\in\tilde{\mathcal{C}}_r} \frac{q_c}{ \sum_{r^\prime\in\mathcal{R}_c}d_c^{r^\prime}\nu_{r^\prime}^*} - 1 \\
& \le & \sum_{r\in\mathcal{R}} \left(\sum_{c^\prime \in \mathcal{C}_{r}}d_{c^\prime}^r q_{c^\prime}\right)
\sum_{c\in{\mathcal{C}}_r} \frac{q_c}{\sum_{r^\prime\in\mathcal{R}_c}d_{c}^{r^\prime}\nu_{r^\prime}^*} - 1 \\
& \le & \sum_{r\in\mathcal{R}} \left(\sum_{c^\prime \in \mathcal{C}_{r}}d_{c^\prime}^r q_{c^\prime}\right)
\sum_{c\in{\mathcal{C}}_r} \frac{d^r_c q_c}{ \sum_{r^\prime\in\mathcal{R}_c}d_{c}^{r^\prime}\nu_{r^\prime}^*} - 1 \\
& \le & \sum_{r\in\mathcal{R}} \left(\sum_{c^\prime \in \mathcal{C}_{r}}d_{c^\prime}^r q_{c^\prime}\right) - 1.
\end{eqnarray*}
The penultimate inequality holds true because $d^r_c \ge 1, \forall c\in\mathcal{C}_r$.
The last inequality comes from the capacity constraint,
by plugging in $\phi_c^{*,1} = \frac{q_c}{\sum_{r\in\mathcal{R}_c} d^r_c\nu^*_r}$
into $\sum_{c\in\mathcal{C}_r} d^r_c \phi_c^{*,1} \le 1$,
we have $\sum_{c\in{\mathcal{C}}_r} \frac{d^r_c q_c}{\sum_{r^\prime\in\mathcal{R}_c}d_c^{r^\prime}\nu_{r^\prime}^*} \le 1$.
Then by swapping the order of summation, we have
\begin{equation*}
\sum_{r\in\mathcal{R}} \left(\sum_{c^\prime \in \mathcal{C}_{r}} d_{c^\prime}^r q_{c^\prime}\right)
= \sum_{c\in\mathcal{C}} q_c \sum_{r\in\mathcal{R}_c}d_c^r = \sum_{c\in\mathcal{C}} q_c D_c.
\end{equation*}
\end{IEEEproof}
\else
\fi

\fi

\section{Elastic Traffic Model}
\label{sec:elastic}
\subsection{System model}
In this section we switch gears to study a scenario 
where the user traffic is elastic, i.e., each user carries
a certain amount of work and leaves the system once
it is finished. 
Specifically, for a class-$c$ user, we assume that 
its service requirement is drawn from an
exponential distribution with mean $\frac{1}{\mu_c}$ independently, and its arrival
follows a Poisson process with intensity $\nu_c$. Then the traffic intensity
associated with user class $c$ is given by $\rho_c = \frac{\nu_c}{\mu_c}$.

Let us first consider a given time instant, when the size of
$\mathcal{U}_c$ and $\mathcal{U}^v$ are given by
$n_c$ and $n^v$ respectively.
Also, for simplicity we assume equal intra-slice weight allocation,
thus
$q_c = \frac{s_{v(c)} n_c}{n^{v(c)}}$.
Substituting $q_c$ into Problem (\ref{eq:scf-def}), 
the $\alpha-$SCS criterion can be rewritten as
follows.
\begin{eqnarray}
\max_{\bm{\phi}} && \sum\limits_{c\in\mathcal{C}} \left( 
\frac{s_{v(c)}n_c}{n^{v(c)}} \right)^\alpha
\frac{(\phi_c)^{1- \alpha}}{1 - \alpha} \label{eq:alpha-fairness}\\
\textrm{such that} && \sum\limits_{c\in\mathcal{C}_r} \phi_c d^r_c \le 1, ~\forall r\in\mathcal{R}.\nonumber
\end{eqnarray}

\subsection{Stability}
Problem (\ref{eq:alpha-fairness}) characterizes the rate allocation
across classes when the numbers of users in the network are fixed.
However, it is natural to study
the evolution of the system when user distributions are random processes. 
Note that while \cite{BoM01} studied the stability condition 
for $\alpha-$fairness when weights are introduced, their weights 
do not depend on the dynamic distribution of users in the network.
By using the fluid system theory established
in \cite{DaM95},\cite{Dai95} and \cite{Dai96}, one can
show that SCS stablizes the system as long as no resource
is overloaded.

\begin{theorem}\label{thm:stability}
Assume that under equal intra-slice weight allocation, 
the rate allocation
is given by Problem (\ref{eq:alpha-fairness}).
Then, when the following effective load conditions are satisfied:
\begin{equation}
\sum_{c\in\mathcal{C}_r} \rho_c d^r_c < 1,~\forall r\in\mathcal{R},
\label{eq:capacity-constraints}
\end{equation}
the network is stable.
\end{theorem}

\textbf{Remark:} 
\Cref{thm:stability} is significant in that the system might become
transient under specific sharing criterion even when Eq. (\ref{eq:capacity-constraints})
is satisfied, e.g., Example 1 in \cite{BoM01}
when strict priorities are designated in favor of
the system throughput.
Moreover, Example 2 in the same literature demonstrates
that even no strict priority is designated,
instability is still possible under Eq. (\ref{eq:capacity-constraints}).
Those examples implies the importance of SCS sharing and associated 
weight allocation schemes.


The result in \cite{BoM01} is under the assumption that
each user has a fixed weight. Thus the \emph{overall} resources committed
to a slice increases with the number of its active users, possibly compromising 
inter-slice protection.
\Cref{thm:stability} shows that even when inter-slice protection
is maintained, SCS can still stablize the system through efficient
utilization.

\iffull
\begin{IEEEproof}
This can be proved by studying the ``fluid system'' associated with
the service discipline proposed. 
Briefly, the ``fluid system'' associated with a queuing system is 
its asymptotic version when the transition frequency is very high and
the change of the queue length in one transition is infinitesimal. 
Such limiting is approached by rescaling the time axis.
The stability of the original queuing system can then be examined by
studying the associated ``fluid system'', see, for example, 
\cite{Dai95}, \cite{Dai96}, and \cite{DaM95}.

According to \cite{Dai95} and \cite{Dai96}, if one can show that such fluid system
gets empty eventually, the associated original queuing system is positive
recurrent.
In view of this result, the outline of the proof is as follows. Firstly
we establish two functions $K(t)$ and $H(t)$ such that $K(t) \ge H(t) \ge 0$,
where $H(t)$ only takes 0 value when all the fluid limits equal to 0.
Then we find a lower bound on the negative drift rate of
$K(t)$ so that we can conclude that $K(t) \rightarrow 0$ eventually.
Therefore $H(t)$, together with all the fluid limits
tend to 0 eventually.

Let us define the vector of users' distribution as
$\mathbf{N}(t) = (N_c(t) : c\in \mathcal{C})$.
Consider the set of ``fluid limits'' defined by:
\begin{equation}
\mathbf{x}(t) = \lim_{\omega \rightarrow \infty} \frac{\mathbf{N}(\omega t)}{\omega}, ~
\textrm{with}~\sum_{c\in\mathcal{C}} N_c(0) = \omega,
\end{equation}
where $\mathbf{x}(t) := (x_c(t) : c\in\mathcal{C})$ is the
vector of fluid limit for each class.
If such limit exists, we have $\sum_{c\in\mathcal{C}} x_c(0) = 1$. 
According to the Lemma 4.2 in \cite{Dai95},
from Strong Law of Large Number one can derive that,
$\mathbf{x}(t)$ is deterministic and
the dynamic of such fluid limits system is actually determined
by the rate allocation problem associated with
the fluid limits.
That is,
$\mathbf{x}(t)$ follows the differential equations:
\begin{equation}
\frac{d}{dt} x_c(t) = \nu_c - \mu_c\tilde{\phi}_c(t),~\textrm{when}~ x_c(t) > 0,
\end{equation}
where $\tilde{\phi}_c(t)$ is the aggregated rate allocated to the fluid limit of
class-$c$, which should be given by the following problem:

\begin{eqnarray}
\max_{\tilde{\bm{\phi}}:=(\tilde{\phi}_c : c\in\mathcal{C})} 
&& \sum\limits_{c\in\mathcal{C}} \left( \frac{s_{v(c)}x_c(t)}{\sum\limits_{c^\prime\in\mathcal{C}^v}x_{c^\prime}(t)} \right)^\alpha
 \frac{\tilde{\phi}_c^{1- \alpha}(t)}{1 - \alpha} \label{eq:fluid-limit-objective}\\
\textrm{such that} && \sum\limits_{c \in \mathcal{C}_r} \tilde{\phi}_c(t) d^r_c \le 1, 
~\forall r\in\mathcal{R}. \nonumber
\end{eqnarray}

Let us assume that $\tilde{\bm{\phi}}(t)$ achieves the maximum of 
Problem (\ref{eq:fluid-limit-objective}).
Then the concavity of the objective function, together with the first-order
optimality condition gives us
$$
G^\prime (\boldsymbol{\zeta})\cdot(\boldsymbol{\zeta} - \boldsymbol{\Lambda}) \le 0,
$$
where $G(\cdot)$ is the objective function of Problem (\ref{eq:fluid-limit-objective}),
for any feasible rate allocation vector $\boldsymbol{\zeta}$. Also note that,
if the capacity constraints Eq. (\ref{eq:capacity-constraints}) are
satisfied by $\bm{\rho}$, there exists $\epsilon > 0$ such that
$(1 + \epsilon) \bm{\rho}$ also satisfies Eq. (\ref{eq:capacity-constraints}). 
Plugging in $(1 + \epsilon)\bm{\rho}$ as $\boldsymbol{\zeta}$
to the above inequality we have:

\begin{eqnarray}
\lefteqn{\sum\limits_{c\in\mathcal{C}} \left( 
\frac{s_{v(c)}x_c(t)}{\sum_{c^\prime\in\mathcal{C}^v}x_{c^\prime}(t)} \right)^\alpha
\rho_c^{-\alpha}(\rho_c - \tilde{\phi}_c(t))} \nonumber\\
& & \le -\epsilon \sum_{c\in\mathcal{C}}
\left( \frac{s_{v(c)}x_c(t)}{\sum_{c^\prime\in\mathcal{C}^v}x_{c^\prime}(t)} \right)^\alpha
 \rho_c^{1 - \alpha}.\label{eq:first-order-condition-expanded}
\end{eqnarray}

If we define function $K(t)$ as
\begin{eqnarray}
K(t) := & \sum\limits_{v\in\mathcal{V}} (s_{v})^\alpha \sum\limits_{c\in\mathcal{C}^v} \int_0^t
\left( \frac{x_c(\tau)}{\sum\limits_{c^\prime\in\mathcal{C}^v}x_{c^\prime}(\tau)} \right)^\alpha
\frac{(\rho_c - \tilde\phi_c(\tau))}{(\rho_c)^\alpha}d\tau \nonumber\\
& + \frac{1}{\bar{\mu}\bar{\rho}^\alpha} \sum\limits_{v\in\mathcal{V}} (s_v)^\alpha
|\mathcal{C}^v|^{-\frac{\alpha^2}{\alpha + 1}}
\|\mathbf{x}^v(0)\|_{\alpha + 1},
\end{eqnarray}
where $\bar{\mu} = \max_c \mu_c$, and $\bar{\rho} = \max_c \rho_c$
are the maximal processing rate and effective load across user types, respectively,
and we define the fluid limit vector of slice $v$ at time $t$ as $\mathbf{x}^v(t) := 
(x_c(t) : c\in\mathcal{C}^v)$, with its L$k-$norm denoted by $\|\mathbf{x}^v(t)\|_k$.
We have that Eq. (\ref{eq:first-order-condition-expanded}) is equivalent to
\begin{equation}
\frac{d}{dt}K(t) \le -\epsilon \sum_{c\in\mathcal{C}}
\left( \frac{s_{v(c)}x_c(t)}{\sum_{c^\prime\in\mathcal{C}^v}x_{c^\prime}(t)} \right)^\alpha
 \rho_c^{1 - \alpha}.
\end{equation}

The right hand side of the above inequality can be bounded by:
\begin{eqnarray*}
\lefteqn{\sum_{c\in\mathcal{C}}
\left( \frac{s_{v(c)}x_c(t)}{\sum_{c^\prime\in\mathcal{C}^v}x_{c^\prime}(t)} \right)^\alpha
 \rho_c^{1 - \alpha}} \\ 
 && \ge s_{min}^\alpha \rho_{bound}^{1-\alpha}
 \sum_v \sum_{c\in\mathcal{C}^v} \left( \frac{x_c(t)}{\sum_{c^\prime\in\mathcal{C}^v}x_{c^\prime}(t)} \right)^\alpha \\
 && \ge s_{min}^\alpha \rho_{bound}^{1-\alpha} \min\{1, (\max_v|\mathcal{C}^v|)^{1-\alpha}\},
\end{eqnarray*}
where $s_{min} = \min_{v} s_v$,
$\rho_{bound}$ takes $\bar{\rho}$ when $\alpha > 1$ and takes $\min_c \rho_c$ when
$0< \alpha < 1$.
The inequality is due to that for each active slice (a slice is said to be active
if $\sum_{c\in\mathcal{C}^v} x_c(t) > 0$), we have two possible cases:
\begin{enumerate}
\item When $0 < \alpha \le 1$, we have
\begin{equation*}
\sum_{c\in\mathcal{C}^v} \left( \frac{x_c(t)}{\sum\limits_{c^\prime\in\mathcal{C}^v} x_{c^\prime}(t)} 
\right)^\alpha
\ge \left(\sum_{c\in\mathcal{C}^v} \frac{x_c(t)}{\sum\limits_{c^\prime\in\mathcal{C}^v} x_{c^\prime}(t)} \right)^\alpha
= 1,
\end{equation*}
due to the concavity of power-$\alpha$. 
\item When $\alpha > 1$, we have
\begin{equation*}
\sum_{c\in\mathcal{C}^v} \left( \frac{x_c(t)}{\sum\limits_{c^\prime\in\mathcal{C}^v} x_{c^\prime}(t)} 
\right)^\alpha
= \frac{\sum\limits_{c\in\mathcal{C}^v} x_c^\alpha(t)}{\left(\sum\limits_{c\in\mathcal{C}^v} x_c(t)\right)^\alpha} \ge |\mathcal{C}^v|^{1-\alpha}.
\end{equation*}
The inequality is
due to that $\|\mathbf{x}^v(t)\|_\alpha |\mathcal{C}^v|^{1 - \frac{1}{\alpha}}
\ge \|\mathbf{x}^v(t)\|_1$ when $\alpha > 1$, see \cite{wiki:norm}.
\end{enumerate}
Thus we found a lower bound for each $v$. By noting that 
there should be 
at least one active user type before the fluid system gets
emptied, we can get the last factor by taking the minimum
across all slices.

Thus, we have
\begin{eqnarray}
\frac{d}{dt}K(t) \le -\epsilon s_{min}^\alpha \rho_{bound}^{1-\alpha} \min\{1, (\max_v|\mathcal{C}^v|)^{1-\alpha}\} \nonumber \\
K(t) \le K(0) -\epsilon s_{min}^\alpha \rho_{bound}^{1-\alpha} \min\{1, (\max_v|\mathcal{C}^v|)^{1-\alpha}\} t.
\end{eqnarray}

In order to find a lower bound of $K(t)$, we observe that
for each slice $v\in\mathcal{V}$ we have
\begin{eqnarray*}
\lefteqn{\sum_{c\in\mathcal{C}_v}\int_0^t
\left( \frac{x_c(\tau)}{\sum_{c^\prime\in\mathcal{C}^v}x_{c^\prime}(\tau)} \right)^\alpha
\rho_c^{-\alpha}(\rho_c - \tilde\phi_c(\tau))d\tau} \\
& \ge & \frac{1}{\bar{\mu}\bar{\rho}^\alpha} \sum_{c\in\mathcal{C}^v} 
\int_0^t \left( \frac{x_c(\tau)}{\sum_{c^\prime\in\mathcal{C}^v}x_{c^\prime}(\tau)} \right)^\alpha
d x_c(\tau),
\end{eqnarray*}
and
\begin{eqnarray*}
\lefteqn{\sum_{c\in\mathcal{C}^v} 
\int_0^t \left( \frac{x_c(\tau)}{\sum_{c\in\mathcal{C}^v}x_c(\tau)} \right)^\alpha
d x_c(\tau)}\\
\lefteqn{\stackrel{y_c(t) := (x_c(t))^{\alpha + 1}}
{=\joinrel=\joinrel=\joinrel=\joinrel=\joinrel=\joinrel=\joinrel=\joinrel=}
\frac{1}{\alpha + 1}\sum_{c\in\mathcal{C}^v} \int_{y_c(0)}^{y_c(t)}
\left( \frac{1}{\sum\limits_{c^\prime\in\mathcal{C}^v} (y_{c^\prime}(\tau))^{\frac{1}{\alpha + 1}}} \right)^\alpha dy_c(\tau)} \\
= && \frac{1}{\alpha + 1}\sum\limits_{c\in\mathcal{C}^v} \int_{y_c(0)}^{y_c(t)}
\left( \left(\sum\limits_{c^\prime\in\mathcal{C}^v} (y_{c^\prime}(\tau))^{\frac{1}{\alpha + 1}}\right)^{\alpha + 1} 
\right)^{-\frac{\alpha}{\alpha + 1}} dy_c(\tau) \\
= &&  \frac{1}{\alpha + 1}\sum\limits_{c\in\mathcal{C}^v} \int_{y_c(0)}^{y_c(t)} 
	\left( \|\mathbf{y}^v(\tau)\|_{\frac{1}{\alpha + 1}} \right)^{-\frac{\alpha}{\alpha + 1}} dy_c(\tau)\\
\ge && \frac{1}{\alpha + 1} \sum\limits_{c\in\mathcal{C}^v}
\int_{y_c(0)}^{y_c(t)} \left(
|\mathcal{C}^v|^\alpha \|\mathbf{y}^v(\tau)\|_1
\right)^{-\frac{\alpha}{\alpha + 1}} dy_c(\tau) \\
= && \frac{|\mathcal{C}^v|^{-\frac{\alpha^2}{\alpha + 1}}}{\alpha + 1}
\int_{\|\mathbf{y}^v(0)\|_1}^{\|\mathbf{y}^v(t)\|_1}
\left( \|\mathbf{y}^v(\tau)\|_1 \right)^{-\frac{\alpha}{\alpha + 1}}
d\left( \|\mathbf{y}^v(\tau)\|_1 \right) \\
= && |\mathcal{C}^v|^{-\frac{\alpha^2}{\alpha + 1}}
\left(
\left(\|\mathbf{y}^v(t)\|_1\right)^{\frac{1}{\alpha + 1}}
-
\left(\|\mathbf{y}^v(0)\|_1\right)^{\frac{1}{\alpha + 1}}
\right) \\
= && |\mathcal{C}^v|^{-\frac{\alpha^2}{\alpha + 1}}
\left( \|\mathbf{x}^v(t)\|_{\alpha + 1}
- \|\mathbf{x}^v(0)\|_{\alpha + 1}\right),
\end{eqnarray*}
where the inequality comes from the relation
between $L1-$norm and $L(\frac{1}{\alpha + 1})-$norm,
and the following equality is by moving the summation
into the integral.
Plugging the above inequality into the definition of $K(t)$, we have 
\begin{eqnarray*}
\lefteqn{K(t) \ge \frac{1}{\bar{\mu}\bar{\rho}^\alpha} 
\sum_{v\in\mathcal{V}} (s_v)^\alpha \sum_{c\in\mathcal{C}^v} \int_0^t 
\left( \frac{x_c(\tau)}{\sum\limits_{c^\prime\in\mathcal{C}^v}x_{c^\prime}(\tau)} \right)^\alpha
d x_c(\tau)} \\
& & + \frac{1}{\bar{\mu}\bar{\rho}^\alpha} \sum_{v\in\mathcal{V}} (s_v)^\alpha
|\mathcal{C}^v|^{-\frac{\alpha^2}{\alpha + 1}}
\|\mathbf{x}^v(0)\|_{\alpha + 1} \\
& \ge & \frac{1}{\bar{\mu}\bar{\rho}^\alpha} \sum_{v\in\mathcal{V}}(s_v)^\alpha
|\mathcal{C}^v|^{-\frac{\alpha^2}{\alpha + 1}}
\left(
\|\mathbf{x}^v(t)\|_{\alpha + 1} - \|\mathbf{x}^v(0)\|_{\alpha + 1}
\right)
\\ 
& & +\frac{1}{\bar{\mu}\bar{\rho}^\alpha} \sum_{v\in\mathcal{V}} (s_v)^\alpha
|\mathcal{C}^v|^{-\frac{\alpha^2}{\alpha + 1}}
\|\mathbf{x}^v(0)\|_{\alpha + 1} \\ 
& = & \frac{1}{\bar{\mu} \bar{\rho}^\alpha} \sum_{v\in\mathcal{V}} (s_v)^\alpha
|\mathcal{C}^v|^{-\frac{\alpha^2}{\alpha + 1}}
\|\mathbf{x}^v(t)\|_{\alpha + 1}.
\end{eqnarray*}

Let us define
\begin{equation}
H(t) := \frac{1}{\bar{\mu} \bar{\rho}^\alpha} \sum_{v\in\mathcal{V}} (s_v)^\alpha
|\mathcal{C}^v|^{-\frac{\alpha^2}{\alpha + 1}}
\|\mathbf{x}^v(t)\|_{\alpha + 1}.
\end{equation}

Thus, we can conclude $K(t) \ge H(t) \ge 0$, where the non-negativity
of $H(t)$ is straightforward, and $H(t) = 0$ only when $x_c(t) = 0, \forall c\in\mathcal{C}$.
Therefore, if we take
$$
T = \frac{F(0)}{\epsilon s_{min}^\alpha \rho_{bound}^{1-\alpha}
\min\{1, (\max_v|\mathcal{C}^v|)^{1-\alpha}\}}, 
$$
$K(t) = H(t) \equiv 0$ when $t \ge T$, implying $x_c(t) \equiv 0$ eventually
for all $c\in\mathcal{C}$.
Thus the system is positive recurrent.
\end{IEEEproof}
\fi

\section{Simulation results}
\label{sec:simulations}
One might think by introducing inter-slice protection,
SCS effectively imposes additional constraints 
to the service discipline, thus is compromised in 
users' performance. However, this needs not to be true,
as we will demonstrate via extensive simulations 
in this section.
We compare the performance of SCS versus 
several benchmarks, including:
\begin{enumerate}
\item Dominant Resource Fairness (DRF) \cite{GZH11}, which is a variation
of weighted maxmin fairness where users' weights are associated with their
resource demands. 
Here to incorporate network slicing,
we use its variation where a user's weight is also associated with
equal intra-slice weight allocation, i.e., 
$w_u = \frac{s_v}{N_v}\cdot 
\delta_u, u\in\mathcal{U}^v$, where $\delta_u$ is 
the dominant share of user $u$
and is given by $\delta_u := \frac{1}{\max_{r\in\mathcal{R}}d^r_c}, u\in\mathcal{U}_c$.
\item (Discriminatory) Processor Sharing (DPS)\cite{AAA06, ChJ07}.
To apply to the multi-resource case, we implement DPS as a variation
of maxmin fairness where user $u$'s weight is $w_u = s_v, u\in\mathcal{U}^v$,
without the notion of per-slice share constraint and inter-slice protection.
\end{enumerate}
Note that because SCS might be hard to scalably compute for general
$\alpha$, we propose the use of $\infty-$SCS,
as a surrogate resource allocation scheme.

In our simulations, we focus on two performance metrics: mean delay and mean throughput.
The delay is defined as the sojourn time of each user, i.e.,
how long it takes for a user to complete service. 
The throughput is defined as the workload divided by
the sojourn time of each user.
The performance of different sharing schemes were compared in a range of
settings, from a simple single resource setting,
to more complex cases where different services/tasks are coupled together 
through shared resources.

\subsection{Single-resource cases}
Since for more complicated network setup, the system performance
(for example, processing rate) is often determined by resource
allocations at certain `bottleneck' resources, we first consider 
single-resource setting. Note that, under such circumstances,
SCS coincides with General Processor Sharing (GPS) \cite{Uit03} 
as well as DRF because all classes
of users $c\in\mathcal{C}$ are associated with the same resource, 
and have the same demand.

\begin{figure}[t!]
\centering
	\subfloat[]{
		\includegraphics[width = 0.37\textwidth]{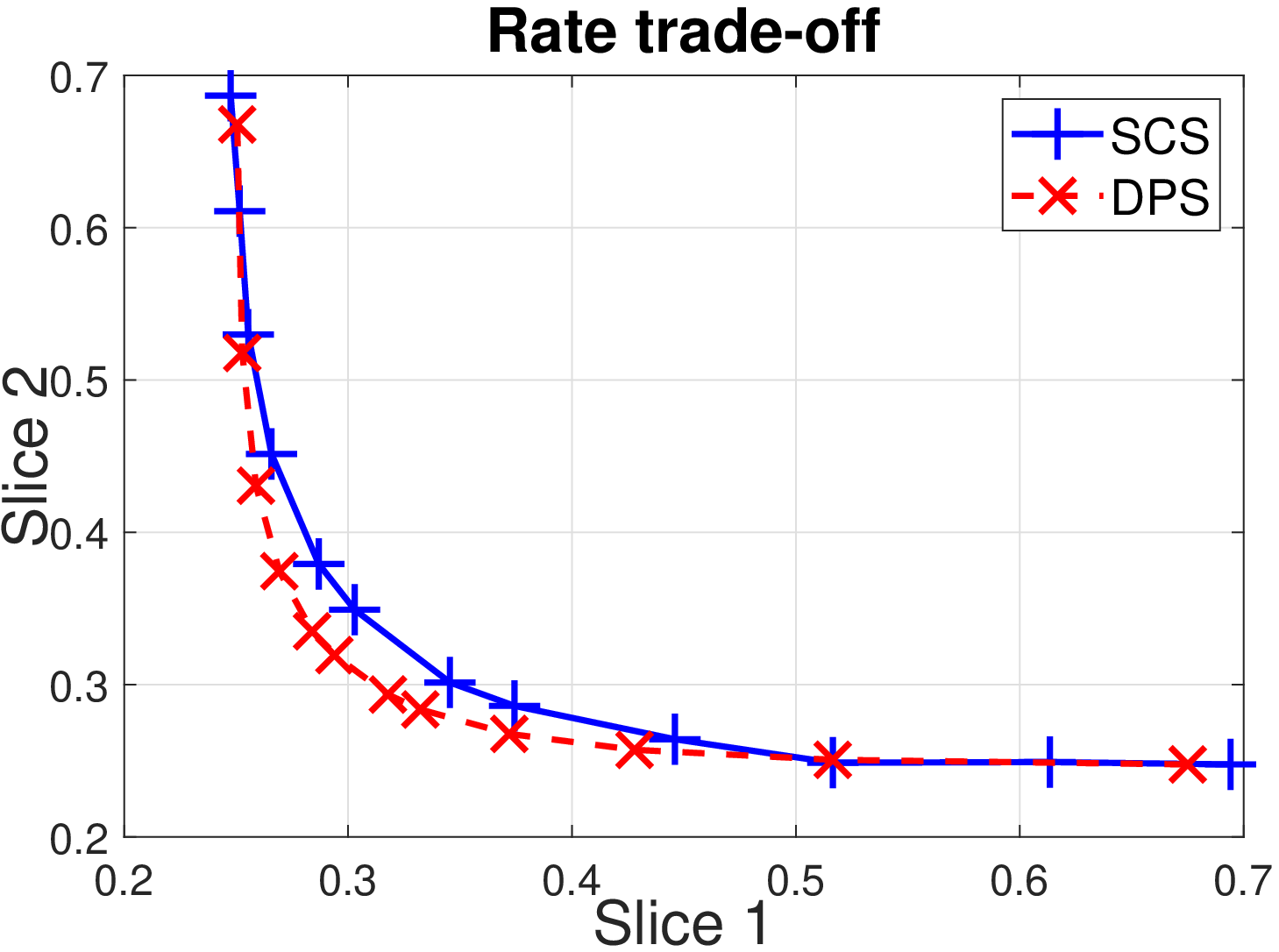}
	}\\
	\subfloat[]{
		\includegraphics[width = 0.37\textwidth]{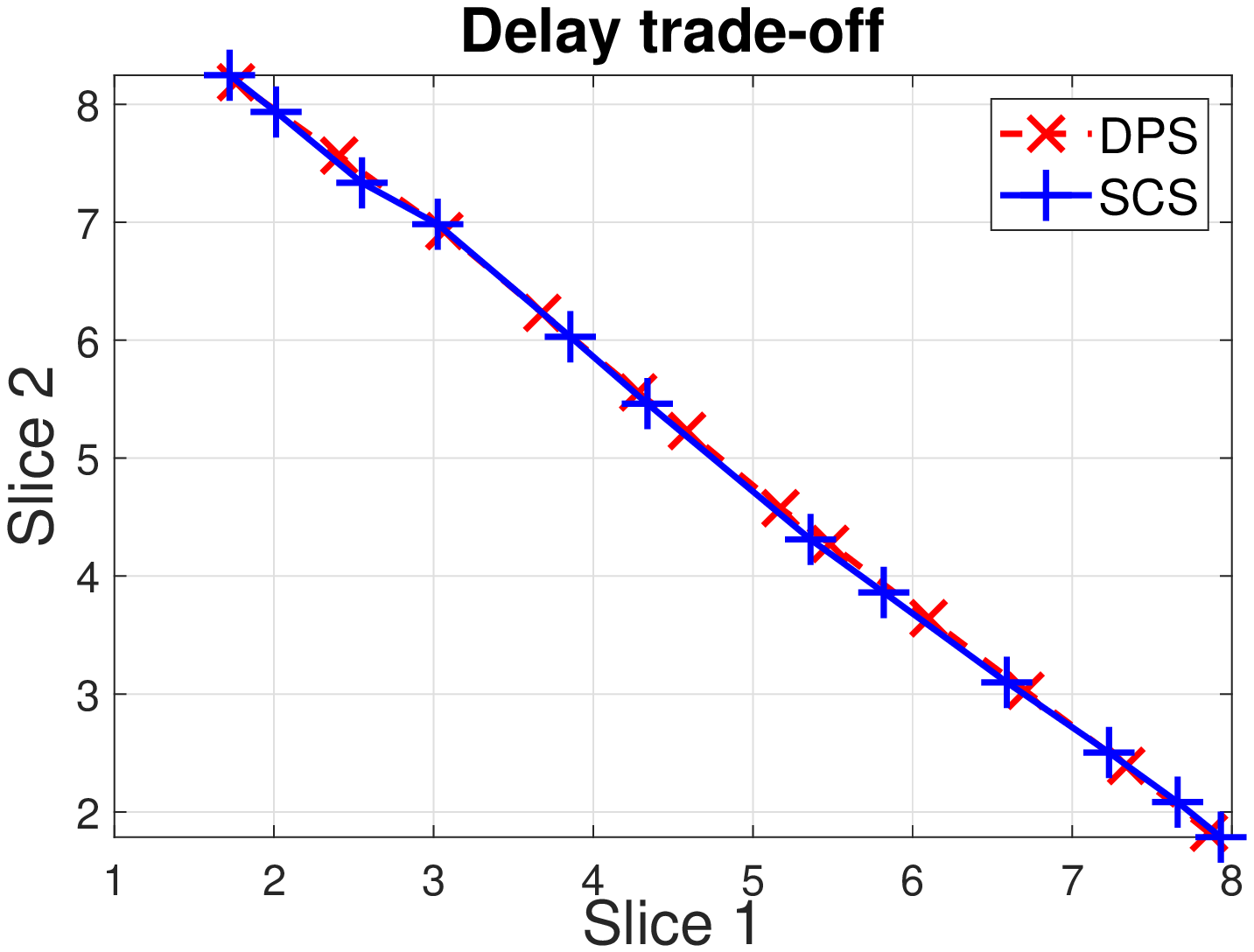}
	}
\caption{Performance trade-offs of single-resource case 
\iffull
under symmetric traffic.
\else
\fi
}
\label{fig:single-symmetric}
\end{figure}

To begin with, we consider a simple scenario where $|\mathcal{V}| = 2$,
and each slice only supports one user class, so in this setting,
a user class corresponds to a slice. Two slices shares one resource,
referred to as Resource $1$
with capacity 1, and $d_1^1 = d_2^1 = 1$.
Their traffic
models are assumed to be symmetric, with mean arrival rates $\nu_1 = \nu_2 = 0.45$
and mean workloads $\frac{1}{\mu_1} = \frac{1}{\mu_2} = 1$.
Their shares, however, are tuned to achieve different performance trade-offs.
The share of Slice 1, $s_1$, ranges from 0.01 to 0.99, while $s_2 = 1 - s_1$.
The achieved mean user perceived delay and 
throughput are illustrated in Fig. \ref{fig:single-symmetric}.
One can see that while the average delays are marginally better under $\infty-$SCS,
$\infty-$SCS
clearly outperforms DPS on the average throughput. For example, 
when two slices have the 
same share $s_1 = s_2 = 0.5$, SCS increases the throughput of users 
on both slice by $\sim$10\%.
\iffull

This phenomenon was widely observed under different traffic assumptions.
For example, when the traffics are asymmetric, with mean arrival rates
$\nu_1 = 0.6, \nu_2 = 0.3$ and mean workload $\frac{1}{\mu_1} = \frac{1}{\mu_2} = 1$,
the results are illustrated in Fig. \ref{fig:single-asymmetric}. Also, for 
symmetric traffics with arrival rates of 0.45 and the workloads are set 
to a constant 1, the results are shown in Fig. \ref{fig:single-md1}.
In general, while the mean delay achieved by SCS is marginally better
than DPS, the mean throughput achieved is improved significantly.
\begin{figure}[t!]
	\centering
	\subfloat[]{
		\includegraphics[width = 0.37\textwidth]{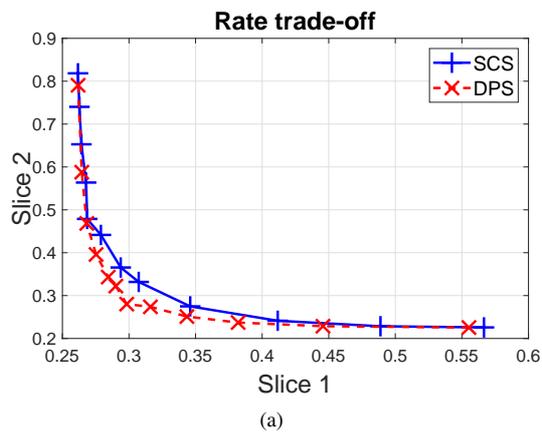}
	}\\
	\subfloat[]{
		\includegraphics[width = 0.37\textwidth]{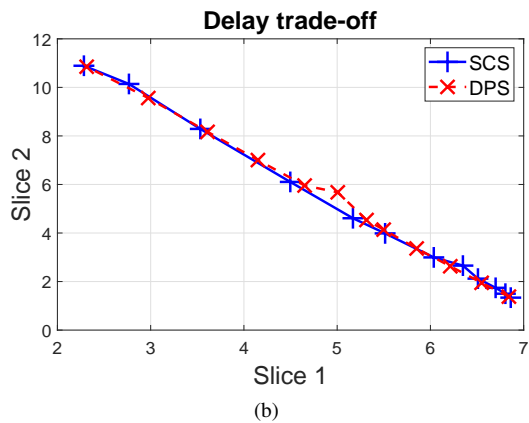}
	}
\caption{Delay and throughput trade-offs of single-resource case under asymmetric traffic.}\label{fig:single-asymmetric}
\end{figure}

\begin{figure}[t!]
	\centering
	\subfloat[]{
		\includegraphics[width = 0.37\textwidth]{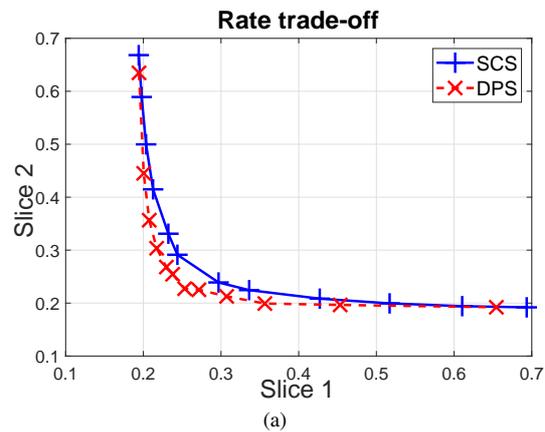}
	}\\
	\subfloat[]{
		\includegraphics[width = 0.37\textwidth]{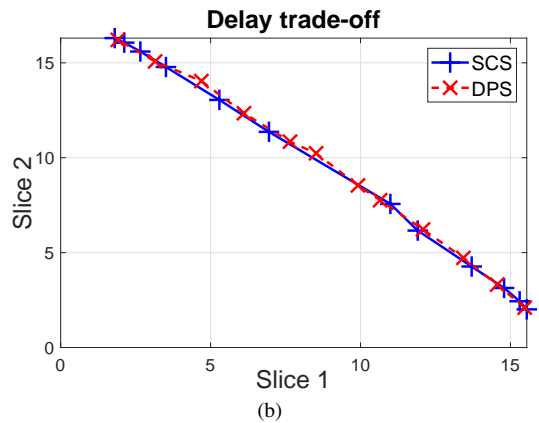}
	}
\caption{Delay and throughput trade-offs of single-resource case under symmetric M/D/1 traffic model.}\label{fig:single-md1}
\end{figure}
\fi

\begin{figure}[t!]
\centering
\includegraphics[width = 0.37\textwidth]{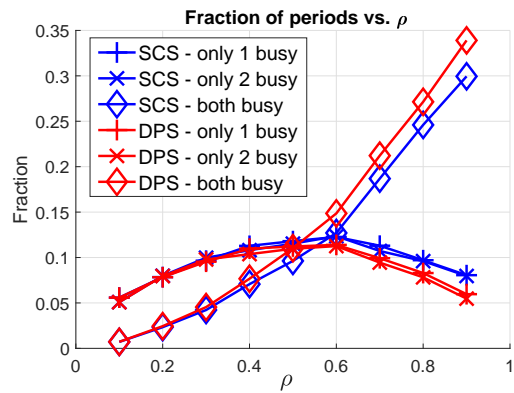}
\caption{Busy period of slice 1 and 2 vs. load intensity.}
\label{fig:busy-period}
\vspace{-1em}
\end{figure}

To explain the somewhat surprising result, 
we conjectured that due to the inter-slice protection built 
into SCS, under stochastic traffic, 
the slice with fewer customers tends to see higher processing rate 
than other sharing criterion, as a result the
customers leave the system faster. Overall, SCS tends to separate the 
busy periods of slices, so that the level of inter-slice contention 
is reduced. We validated our conjecture by measuring the busy 
period under the symmetric traffic pattern, where the arrival rates of 
both slices are the same, and are tuned from 0.05 to 0.45, with $s_1 = s_2 = 0.5$.
Other parameters are the same as in the setting in 
Fig. \ref{fig:single-symmetric}. We plot the fraction of times 
when there is only one busy slice and both slices are busy, vs. 
the effective traffic intensity $\rho = \frac{\nu_1}{\mu_1} + \frac{\nu_2}{\mu_2}$ 
in Fig. \ref{fig:busy-period}. One can see that, for both SCS and DPS, 
the time fraction when both slices are busy increases with $\rho$,
and that when only one slice is busy first increases when $\rho$ 
is low due to underutilization, but decreases when $\rho$ is 
high because the inter-slice contention becomes inevitable. 
However the time fraction when both are busy is always smaller 
under SCS than that under DPS.

\subsection{Multiple-resource cases}
We also test the performance of SCS under a more complex setting 
where a simple cellular networks with both fronthaul and backhaul 
resources are simulated.

\begin{figure}[t!]
\centering
\includegraphics[width = 0.45\textwidth]{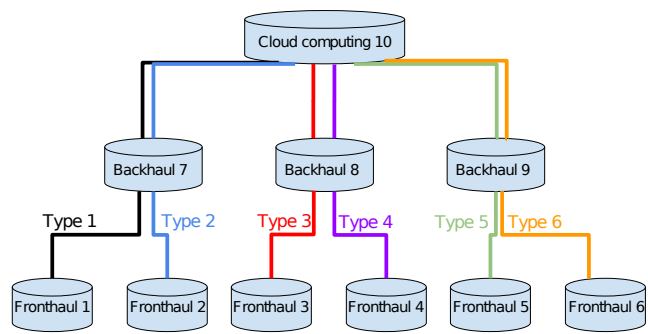}
\caption{Association between user classes and resources.}
\label{fig:multi-resource-setup}
\end{figure}

\begin{figure}[t!]
	\centering
	\subfloat[]{
		\includegraphics[width = 0.37\textwidth]{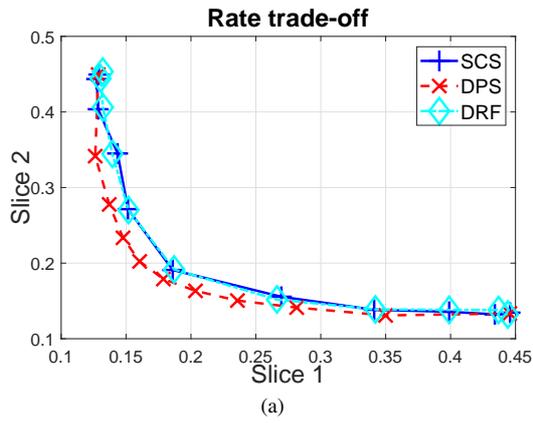}
	}\\
	\subfloat[]{
		\includegraphics[width = 0.37\textwidth]{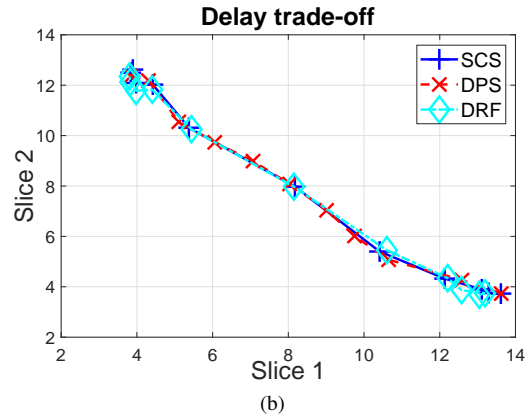}
	}
\caption{Performance trade-offs of multi-resource case.}
\label{fig:multi-constrained-drf}
\end{figure}

\begin{figure}[t!]
	\centering
	\subfloat[]{
		\includegraphics[width = 0.37\textwidth]{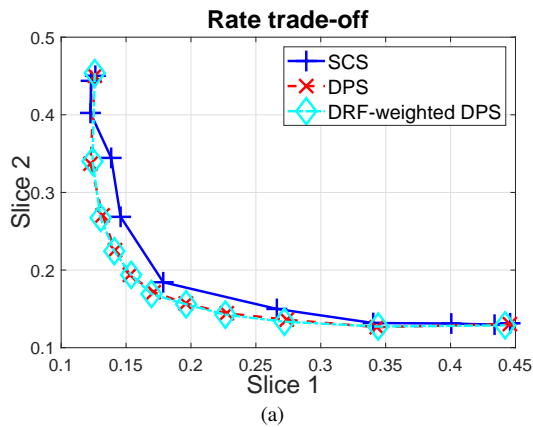}
	}\\
	\subfloat[]{
		\includegraphics[width = 0.37\textwidth]{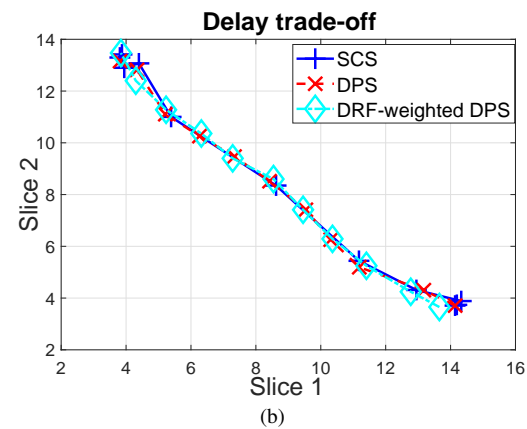}
	}
\caption{Performance trade-offs with DRF-weighted DPS.}
\label{fig:multi-constrained-drfdps}
\vspace{-1em}
\end{figure}

Let us consider a setting with 6 fronthaul resources, 3 backhaul resources, and
a cloud computing resource. This system supports two slices, each 
containing 3 user classes. Slice 1 includes Classes 1,2 and 3, while
Slice 2 includes Classes 4, 5 and 6. The association between user 
classes and resources is demonstrated in Fig.~\ref{fig:multi-resource-setup},
and the demand vectors, as well as the arrival rates and mean workloads,
are given in Table \ref{tab:sim-setup-multi}.
Slice 1's share is ranged from 0.1 to 0.9, 
while $s_2 = 1-s_1$. The achieved performance trade-offs 
under different sharing criteria 
are illustrated in Fig. \ref{fig:multi-constrained-drf}. One can see that 
both SCS and DRF outperform DPS in throughput, with similar mean delays
under all 3 criteria. 
Similar results are observed in a range of settings
with different traffic patterns and resource demands.
Moreover,
in Fig. \ref{fig:multi-constrained-drfdps}, we adjust the weighting schemes used
in DRF by voiding SCWA.
Instead, $w_u = s_v\delta_u, u\in\mathcal{U}^v$, and the resources 
are provisioned according to DPS with weight $w_u$.
The results show that without SCWA, 
DRF is similar to DPS in both throughput and delay. 
Therefore, we can conclude that SCWA is the root cause of the desirable performance, 
and SCS can even
improve the system performance while providing inter-slice protection.

\begin{table}[t!]
\resizebox{0.48\textwidth}{!}{
\begin{tabular}{|l||l|l|l|l|l|l|}
\hline
User & \multirow{2}{*}{Demand vector} & Mean          & Arrival \\ 
class&                                & workload      & rate    \\ \hline\hline
Class 1 & $(\frac56,0,0,0,0,0,0.5,0,0,0.217)$ & 1 & 0.7  \\ \hline
Class 2 & $(0,\frac56,0,0,0,0,0.5,0,0,0.217)$ & 1 & 0.7  \\ \hline
Class 3 & $(0,0,1,0,0,0,0,0.625,0,0.217)$       & 1 & 0.7  \\ \hline
Class 4 & $(0,0,0,1,0,0,0,0.625,0,0.217)$       & 1 & 0.7  \\ \hline
Class 5 & $(0,0,0,0,1,0,0,0,0.625,0.217)$ & 1 & 0.7  \\ \hline
Class 6 & $(0,0,0,0,0,1,0,0,0.625,0.217)$ & 1 & 0.7  \\ \hline
\end{tabular}
}
\caption{Example resource allocation}
\label{tab:sim-setup-multi}
\vspace{-1em}
\end{table}

\section{Conclusions and future work}
This paper has explored a novel approach to resource allocation 
for network slicing 
of distributed resources--SCS,
providing inter-slice protection, load-driven elasticity and desirable
performance at the same time.
SCS can be viewed as a key to enabling low-complexity
performance management in network slicing, by exposing
\emph{network shares} to slice operators/tenants, 
as a high-level resource management interface. 
This approach can be further extended in two directions:
i) if slices have highly imbalanced spatial user distributions,
it might be useful to let slices specify different
shares across different pools of resources, e.g.,
regions corresponding to downtown, stadium and/or
rural area, see, e.g., \cite{pablo18}; and
ii) slices may wish to request different shares across
types of resources, e.g., a slice may specify a higher
share of computational resource pool than that of the 
communicational resources. For example, mobile cloud gaming
is computation intensive, thus the operator might
want to reserve more computing resources than connectivity.
Finally, the share abstraction provides a simple
parametric ``crude'' model for slice-level resource
allocation which needs to interact with an intra-slice performance
management strategy.


\bibliographystyle{abbrv}
\bibliography{sharing}

\end{document}